\documentclass{llncs}

\pdfoutput=1
\usepackage{makeidx}  

\usepackage{amssymb,amsmath,amsfonts}
\usepackage{pifont}
\usepackage{graphicx}
\usepackage{latexsym}
\usepackage{todonotes}
\usepackage{xspace}
\usepackage{tikz,pgf}
\usepackage{paralist}
\usepackage{xcolor}
\usepackage{wasysym}
\usepackage{wrapfig}

\usetikzlibrary{automata,calc,arrows,shapes,backgrounds}

\definecolor{ly}{RGB}{250,250,200}
\definecolor{lg}{RGB}{200,250,200}

\def\Nset{\mathbb{N}}

\def\ra{\rightarrow}

\newcommand{\thmref}[1]{Theorem~\ref{#1}}
\newcommand{\propref}[1]{Proposition~\ref{#1}}

\newcommand{\lemref}[1]{Lemma~\ref{#1}}
\newcommand{\figref}[1]{Figure~\ref{#1}}

\newcommand{\IGNORE}[1]{}

\newcommand{\dist}{\mathsf{Dist}}
\newcommand{\adist}{\mu}

\newcommand{\support}{\mathsf{support}}

\newcommand{\dirac}[1]{\{ {#1} \mapsto 1 \}}

\newcommand{\states}{S}
\newcommand{\astate}{s}

\newcommand{\last}{\mathit{last}}

\newcommand{\set}[1]{\{{#1}\}}
\newcommand{\ldot}{\, . \,}

\newcommand{\lcrc}{[\cdot,\cdot]}
\newcommand{\lorc}{(\cdot,\cdot]}
\newcommand{\lcro}{[\cdot,\cdot)}
\newcommand{\loro}{(\cdot,\cdot)}
\newcommand{\lnz}{\langle +,\cdot \rangle}
\newcommand{\lcz}{[0,\cdot \rangle}
\newcommand{\loz}{(0,\cdot \rangle}
\newcommand{\rc}{\langle \cdot, \cdot]}
\newcommand{\lz}{\langle 0, \cdot \rangle}

\newcommand{\intervals}{\mathcal{I}}
\newcommand{\anint}{I}
\newcommand{\lep}[1]{\mathsf{left}({#1})}
\newcommand{\rep}[1]{\mathsf{right}({#1})}
\newcommand{\ilcrc}{\intervals^{\lcrc}}
\newcommand{\ilorc}{\intervals^{\lorc}}
\newcommand{\ilcro}{\intervals^{\lcro}}
\newcommand{\iloro}{\intervals^{\loro}}
\newcommand{\ilnz}{\intervals^{\lnz}}
\newcommand{\ilcz}{\intervals^{\lcz}}
\newcommand{\iloz}{\intervals^{\loz}}

\newcommand{\ilz}{\intervals^{\lz}}

\newcommand{\adtmc}{\mathcal{D}}
\newcommand{\dstates}{S}
\newcommand{\dtrans}{\mathbf{P}}
\newcommand{\dpaths}[1]{\mathit{Paths}^{#1}}
\newcommand{\dfinpaths}[1]{\mathit{Paths}^{#1}_*}
\newcommand{\dprob}[2]{\mathrm{Pr}^{#1}_{#2}}
\newcommand{\adtmcparaone}[1]{\adtmc^{#1}}
\newcommand{\adtmcparatwo}[2]{\adtmc^{#1}_{#2}}
\newcommand{\dtranspara}[1]{\dtrans_{#1}}
\newcommand{\afolddtmcpara}[1]{\tilde{\adtmc}^{#1}}
\newcommand{\folddtrans}{\tilde{\dtrans}}

\newcommand{\apath}{\rho}

\newcommand{\anmdp}{\mathcal{M}}
\newcommand{\mstates}{S}
\newcommand{\mtrans}{\Delta}
\newcommand{\mpaths}[1]{\mathit{Paths}^{#1}}
\newcommand{\mfinpaths}[1]{\mathit{Paths}^{#1}_*}
\newcommand{\afinpath}{r}
\newcommand{\afinpathpref}[1]{r_{#1}}
\newcommand{\mprob}[2]{\mathrm{Pr}^{#1}_{#2}}
\newcommand{\asched}{\sigma}
\newcommand{\scheds}[1]{\Sigma^{#1}}
\newcommand{\memlessscheds}[1]{\Sigma^{#1}_\mathrm{m}}

\newcommand{\anoimc}{\mathcal{O}}
\newcommand{\ostates}{S}
\newcommand{\otrans}{\delta}

\newcommand{\edges}{E}
\newcommand{\edgespara}[1]{\edges^{#1}}
\newcommand{\anedge}{e}

\newcommand{\apa}[1]{\mathbf{a}}
\newcommand{\apatwo}[2]{\mathbf{a}_{#1}^{#2}}

\newcommand{\uident}{\mathrm{U}}
\newcommand{\iident}{\mathrm{I}}

\newcommand{\umc}[1]{[{#1}]_\uident}
\newcommand{\imdp}[1]{[{#1}]_\iident}

\newcommand{\targetset}{T}
\newcommand{\reachpaths}[1]{\mathsf{Reach}({#1})}

\newcommand{\eqset}[2]{\ostates^{#1,#2}_{\exists}}
\newcommand{\aqset}[2]{\ostates^{#1,#2}_{\forall}}

\newcommand{\edgesfrom}[1]{\edges({#1})}
\newcommand{\edgesubset}{B}
\newcommand{\edgesubsetpara}[1]{\edgesubset_{#1}}
\newcommand{\edgesubsetparatwo}[2]{\edgesubset^{#1}_{#2}}

\newcommand{\validsetsfrom}[1]{\mathit{Valid}({#1})}
\newcommand{\validsets}{\mathit{Valid}}
\newcommand{\waf}{w}
\newcommand{\possass}[1]{\mathit{ValidAssign}({#1})}

\newcommand{\qa}[2]{[{#1}]_{#2}}
\newcommand{\qatrans}[1]{\mtrans_{#1}}

\newcommand{\gpaths}[1]{\mathit{Paths}^{#1}}
\newcommand{\gfinpaths}[1]{\mathit{Paths}^{#1}_*}

\newcommand{\astateset}{X}
\newcommand{\anotherstateset}{Y}
\newcommand{\cpre}{\mathsf{CPre}}
\newcommand{\dualcpre}{\overline{\mathsf{CPre}}}
\newcommand{\apre}{\mathsf{APre}}

\newcommand{\aschedpara}[1]{\asched_{#1}}

\newcommand{\abscc}{U}
\newcommand{\bsccset}{\mathcal{U}}

\newcommand{\vsa}[1]{\mathfrak{f}({#1})}

\newcommand{\asv}[1]{\mathfrak{g}({#1})}
\newcommand{\asvnopara}{\mathfrak{g}}

\newcommand{\ecstates}{C}
\newcommand{\ecdists}{D}
\newcommand{\ecs}{\mathcal{E}}
\newcommand{\infalong}[1]{\mathit{inf}({#1})}
\newcommand{\sa}[1]{\mathit{sa}{#1}}

\newcommand{\edgesfromthree}[3]{\edges^{#1}({#2},{#3})}
\newcommand{\edgesfromtwo}[2]{\edges({#1},{#2})}

\newcommand{\ilecstates}{C}
\newcommand{\ilecs}{\mathfrak{I}}

\newcommand{\infstatesalong}[1]{\mathit{infst}({#1})}

\newcommand{\startindex}{\mathfrak{i}}

\newcommand{\boundedreachpaths}[2]{\mathit{Reach}^{\leq {#1}}({#2})}

\newcommand{\lbs}[1]{\kappa_{#1}}
\newcommand{\lbsilec}{\overline{\kappa}}
\newcommand{\lbl}[1]{\lambda_{#1}}
\newcommand{\lblilec}{\overline{\lambda}}

\newcommand{\statesettwo}[2]{U_{\neg {#2}}}
\newcommand{\ilecstatespara}[1]{\ilecstates_{#1}}

\begin{document}
\title{Qualitative Reachability for Open Interval Markov Chains}
\titlerunning{Qualitative Reachability for Open Interval Markov Chains}  
%
\author{
Jeremy Sproston
}
\authorrunning{Jeremy Sproston} 
%
\tocauthor{
Jeremy Sproston
}
\institute{
Dipartimento di Informatica,
University of Turin, Italy
}

\maketitle              

\begin{abstract}
Interval Markov chains extend classical Markov chains with the possibility to describe transition probabilities using intervals, rather than exact values. While the standard formulation of interval Markov chains features closed intervals, previous work has considered also open interval Markov chains, in which the intervals can also be open or half-open. In this paper we focus on qualitative reachability problems for open interval Markov chains, which consider whether the optimal (maximum or minimum) probability with which a certain set of states can be reached is equal to 0 or 1. We present polynomial-time algorithms for these problems for both of the standard semantics of interval Markov chains. Our methods do not rely on the closure of open intervals, in contrast to previous approaches for open interval Markov chains, and can characterise situations in which probability 0 or 1 can be attained not exactly but arbitrarily closely.
\end{abstract}
%

\section{Introduction}\label{sec:intro}
The development of modern computer systems can benefit substantially from a verification phase,
in which a formal model of the system is exhaustively verified 
in order to identify undesirable errors or inefficiencies.
In this paper we consider the verification of probabilistic systems,
in which state-to-state transitions are accompanied by probabilities that specify the relative likelihood
with which the transitions occur, using model-checking techniques;
see \cite{BK08,FKNP11,BdAFK18} for general overviews of this field.
One drawback of classical formalisms for probabilistic systems is that they typically require the specification
of exact probability values for transitions:
in practice, it is likely that such precise information concerning 
the probability of system behaviour is not available.
A solution to this problem is to associate intervals of probabilities with transitions,
rather than exact probability values,
leading to \emph{interval Markov chains} (IMCs) or \emph{interval Markov decision processes}.
IMCs have been studied in \cite{JL91,KU02},
and considered in the \emph{qualitative} and \emph{quantitative} 
model-checking context in \cite{SVA06,CSH08,CHK13}.
Qualitative model checking concerns whether a property is satisfied by the system model with probability
(equal to or strictly greater than) $0$ or (equal to or strictly less than) $1$,
whereas quantitative model checking considers whether a property is satisfied
with probability above or below some threshold in the interval $[0,1]$,
and generally involves the computation of the probability of property satisfaction,
which is then compared to the aforementioned threshold.


\begin{figure}[t]
\centering
\begin{minipage}{.4\textwidth}

{

\centering
\scriptsize

\begin{tikzpicture}[->,>=stealth',shorten >=1pt,auto, thin] 

  \tikzstyle{state}=[draw=black, text=black, shape=circle, inner sep=3pt, outer sep=0pt, circle, rounded corners, fill=ly] 
  \tikzstyle{final_state}=[draw=black, text=black, shape=circle, inner sep=3pt, outer sep=0pt, circle, rounded corners, fill= lg] 

	\node[state, node distance=1cm](A){$\astate_0$};

	\node[state, node distance=2cm](B)[right of=A]{$\astate_1$};
		
	\path[rounded corners]

	(A)
		edge [loop above, above] node {$(0,1)$} (A)

	(A)
		edge [above] node {$(0,1)$} (B)

	(B)
		edge [loop above, above] node {$[1,1]$} (B);

\end{tikzpicture}

}

\caption{An open IMC $\anoimc_1$.}
\label{fig:open_example}
\end{minipage}%
\begin{minipage}{.5\textwidth}
{

\centering
\scriptsize

\begin{tikzpicture}[->,>=stealth',shorten >=1pt,auto, thin] 

  \tikzstyle{state}=[draw=black, text=black, shape=circle, inner sep=3pt, outer sep=0pt, circle, rounded corners, fill=ly] 
  \tikzstyle{final_state}=[draw=black, text=black, shape=circle, inner sep=3pt, outer sep=0pt, circle, rounded corners, fill= lg] 

	\node[state, node distance=1cm](A){$\astate_0$};

	\node[state, node distance=2cm](B)[right of=A]{$\astate_1$};

	\node[state, node distance=2cm](C)[right of=B]{$\astate_2$};

	\path[rounded corners]
	(A)
		edge [loop above, above] node {$(0,0.6)$} (A)

	(A)
		edge [above, bend right, bend angle=60] node {$(0.5,1)$} (B)

	(B)
		edge [loop above] node {$[0,0.5]$} (B)

	(B)
		edge [above, bend right, bend angle=60] node {$[0.6,0.8]$} (A)

	(B)
		edge [above] node {$(0,0.2]$} (C)

	(C)
		edge [loop above, above] node {$[1,1]$} (C);

\end{tikzpicture}

}

\caption{An open IMC $\anoimc_2$.}
\label{fig:general_example}
\end{minipage}

\end{figure}


In \cite{SVA06,CSH08,CHK13}, the intervals associated with transitions are \emph{closed}.
This limitation was adressed in \cite{CK15},
which considered the possibility of utilising \emph{open} (and half-open) intervals,
in addition to closed intervals.
Example of such open IMCs are shown in \figref{fig:open_example} and \figref{fig:general_example}.
In \cite{CK15}, it was shown that the probability of the satisfaction of a property
in an open IMC
can be approximated arbitrarily closely by a standard, closed IMC
obtained by changing all (half-)open intervals featured in the model
to closed intervals with the same endpoints.
However, although the issue of the determining the existence of exact solutions is mentioned in \cite{CK15},
closing the intervals can involve the loss of information concerning exact solutions.
Take, for example, the open IMC in Figure \ref{fig:open_example}:
changing the intervals from $(0,1)$ to $[0,1]$ on both of the transitions
means that the minimum probability of reaching the state $\astate_1$ after starting in state $\astate_0$
becomes $0$,
whereas the probability of reaching $\astate_1$ from $\astate_0$ is strictly greater than $0$
for all ways of assigning probabilities to the transitions in the original IMC.

In this paper we propose verification methods for 
qualitative reachability properties of open IMCs.
We consider both of the standard semantics for IMCs.
The uncertain Markov chain (UMC) semantics associated with an IMC comprises an 
infinite number of standard Markov chains,
each corresponding to a certain choice of probability for each transition.
In contrast, the interval Markov decision process (IMDP) semantics associates 
a single Markov decision process (MDP)
with the IMC,
where from each state there is available an uncountable number of distributions,
each corresponding to one assignment of probabilities belonging to the intervals of the transitions
leaving that state.
The key difference between the two semantics can be summarised by 
considering the behaviour from a particular state of the IMC:
in the UMC semantics, the same probability distribution over outgoing transitions
must always be used from the state,
whereas in the IMDP semantics
the outgoing probability distribution may change for each visit to the state.
We show that we can obtain exact (not approximate) solutions
for both semantics
in polynomial time in the size of the open IMC.

For the UMC semantics,
and for three of the four classes of qualitative reachability problem 
in the IMDP semantics,
the algorithms presented are inspired by methods for finite MDPs.
In the case of the IMDP semantics,
these algorithms rely on the fact that retaining the memory of 
previous choices along the behaviour of an IMC is not necessary.
A direct method for the construction of a finite MDP that represents an IMC
and which can be used for the verification of qualitative properties is the following:
the set of states of the finite MDP equals that of the IMC and,
for each state $\astate$ and each set $\astateset$ of states,
there exists a single distribution from $\astate$ in the finite MDP 
that assigns positive probability to each state in $\astateset$ 
if and only if
there exists at least one probability assignment for transitions in the IMC
that assigns positive probability to each transition from $\astate$ with target state in $\astateset$.
Intuitively,
a distribution associated with $\astate$ and $\astateset$ in the finite MDP 
can be regarded as the representative distribution of all probability assignments of the IMC
that assign positive probability to the transitions from $\astate$ to states in $\astateset$.
However, such a finite MDP construction 
does not yield polynomial-time algorithms in the size of the open IMC,
because the presence of transitions having zero as their left endpoint
can result in an exponential number of distributions in the number of IMC transitions.
In our methods, apart from considering issues concerning the difference between closed and open intervals
and the subsequent implications for qualitative reachability problems,
we avoid such an exponential blow up.
In particular, we show how the predecessor operations used by some qualitative reachability algorithms
for MDPs
can be applied directly on the open IMC.

The fourth class of reachability problem in the IMDP semantics 
concerns determining whether the probability of reaching a certain set of states
from the current state is equal to $1$ for all schedulers,
where a scheduler chooses an outgoing probability distribution from a state
on the basis of the choices made so far.
For this class of problem,
retaining memory of previous choices can be important for showing that the problem is \emph{not} satisfied,
i.e., that there exists a scheduler such that the reachability probability is strictly less than $1$.
As an example, we can take the open IMC in Figure \ref{fig:open_example}.
Consider the memoryful scheduler that 
assigns probability $\frac{1}{2^i}$ to the $i$-th attempt to take a transition from $\astate_0$ to $\astate_1$,
meaning that the overall probability of reaching $\astate_1$ when starting in $\astate_0$
under this scheduler
is $\frac{1}{2} + \frac{1}{2}(\frac{1}{4} + \frac{3}{4}( \frac{1}{8} + \cdots )) < 1$.
Instead a memoryless scheduler will reach $\astate_1$ with probability 1:
for any $\lambda \in (0,1)$ representing 
the (constant) probability of taking the transition from $\astate_0$ to $\astate_1$,
the overall probability of reaching $\astate_1$ is 
$\lim_{k \ra \infty} 1-(1-\lambda)^k = 1$.
Hence our results for this class of reachability problem take 
the inadequacy of memoryless schedulers into account;
indeed, while the algorithms presented for all other classes of problems
(and all problems for the UMC semantics)
proceed in a manner similar to that introduced in the literature for finite MDPs,
for this class we present an \emph{ad hoc} algorithm, 
based on an adaptation of the classical notion of end components \cite{deAlf97}.

After introducing open IMCs in Section \ref{sec:iomcs},
the algorithms for the UMC semantics and the IMDP semantics
are presented in Section \ref{sec:umc_algos} and Section \ref{sec:imdp_algos}, respectively.
The proofs of the results can be found in the appendix.

{\sl Related work.}
Model checking of qualitative properties of Markov chains (see, for example, \cite{Var85,CY95})
relies on the fact that transition probability values are fixed throughout the behaviour of the system,
and does not require that exact probability values are taken into account during analysis. 
The majority of work on model checking for IMCs 
considers the more general quantitative problems:
\cite{SVA06,CSH08} present algorithms utilising a finite MDP construction
based on encoding within distributions available from a state 
the extremal probabilities allowed from that state (known as the state's basic feasible solutions).
Such a construction results in an exponential blow up,
which is also not avoided in \cite{CSH08} for qualitative properties
(when transitions can have $0$ as their left endpoint).
\cite{CHK13,PLSS13} improve on these results to present polynomial-time algorithms
for reachability problems based on linear or convex programming.
The paper \cite{HM18} includes polynomial-time methods for 
computing (maximal) end components,
and for computing a single step of value iteration, for interval MDPs.
We note that IMCs are a special case of constraint Markov chains \cite{CDLLPW11},
and that the UMC semantics of IMCs corresponds to a special case of 
parametric Markov chains \cite{Daw04,LMT07}.
As far as we are aware, only \cite{CK15} considers open IMCs.

\section{Open Interval Markov Chains}\label{sec:iomcs}

\subsubsection{Preliminaries.}
%
%
A \emph{(probability) distribution} over a finite set $Q$ is a function $\adist: Q \ra [0,1]$
such that $\sum_{q \in Q} \adist(q) = 1$.
Let $\dist(Q)$ be the set of distributions over $Q$.
We use $\support(\adist) = \set{ q \in Q \mid \adist(q)>0 }$
to denote the \emph{support set} of $\adist$,
i.e., the set of elements assigned positive probability by $\adist$,
and use $\dirac{q}$ to denote the distribution that assigns probability $1$ to the single element $q$.
Given a binary function $f: Q \times Q \ra [0,1]$ and element $q \in Q$,
we denote by $f(q,\cdot): Q \ra [0,1]$ the unary function 
such that $f(q,\cdot)(q')=f(q,q')$ for each $q' \in Q$.
We let $\intervals$ denote the set of (open, half-open or closed) intervals that are subsets of $[0,1]$
and that have rational-numbered endpoints.
Given an interval $\anint \in \intervals$,
we let $\lep{\anint}$ (respectively, $\rep{\anint}$) be the left (respectively, right) endpoint of $\anint$.
The set of closed (respectively, left-open, right-closed; left-closed, right-open; open) intervals in $\intervals$ 
is denoted by $\ilcrc$ (respectively, $\ilorc$; $\ilcro$; $\iloro$).
Hence we have $\intervals = \ilcrc \cup \ilorc \cup \ilcro \cup \iloro$.
Furthermore, we let $\ilnz$ (respectively, $\ilcz$; $\iloz$) be the set of intervals in $\intervals$ 
such that the left endpoint is positive 
(respectively, left-closed intervals with the left endpoint equal to $0$;
left-open intervals with the left endpoint equal to $0$).
Finally, let $\ilz = \ilcz \cup \iloz$ be the set of intervals in $\intervals$ with left endpoint equal to zero.


A \emph{discrete-time Markov chain} (DTMC) $\adtmc$ is a pair $(\dstates,\dtrans)$
where $\dstates$ is a set of \emph{states},
and $\dtrans: \dstates \times \dstates \ra [0,1]$ is a \emph{transition probability matrix},
such that, for each state $\astate \in \dstates$, 
we have $\sum_{\astate' \in \dstates} \dtrans(\astate,\astate') = 1$.
Note that $\dtrans(\astate,\cdot)$ is a distribution, for each state $\astate \in \dstates$.
A \emph{path} of DTMC $\adtmc$ is a sequence $\astate_0 \astate_1 \cdots$
such that $\dtrans(\astate_i,\astate_{i+1})>0$ for all $i \geq 0$.
Given a path $\apath = \astate_0 \astate_1 \cdots$ and $i \geq 0$,
we let $\apath(i) = \astate_i$ be the $(i+1)$-th state along $\apath$.
The set of paths of $\adtmc$ starting in state $\astate \in \dstates$ is denoted by $\dpaths{\adtmc}(\astate)$.
In the standard manner (see, for example, \cite{BK08,FKNP11}),
given a state $\astate \in \dstates$,
we can define a probability measure $\dprob{\adtmc}{\astate}$ over $\dpaths{\adtmc}(\astate)$.


A \emph{Markov decision process} (MDP) $\anmdp$ is a pair $(\mstates,\mtrans)$
where $\mstates$ is a finite set of \emph{states}
and $\mtrans: \states \ra 2^{\dist(\mstates)}$
is a \emph{transition function} such that $\mtrans(\astate) \neq \emptyset$ for all $\astate \in \mstates$.
We say that an MDP is \emph{finite} if $\mtrans(\astate)$ is finite for all $\astate \in \mstates$.

A(n infinite) path of an MDP $\anmdp$ is a sequence $\astate_0 \adist_0 \astate_1 \adist_1 \cdots$
such that $\adist_i \in \mtrans(\astate_i)$ and $\adist_i(\astate_{i+1})>0$
for all $i \geq 0$.
Given a path $\apath = \astate_0 \adist_0 \astate_1 \adist_1 \cdots$ and $i \geq 0$,
we let $\apath(i) = \astate_i$ be the $(i+1)$-th state along $\apath$.
A finite path is a sequence $\afinpath = \astate_0 \adist_0 \astate_1 \adist_1 \cdots \adist_{\adist_{n-1}} \astate_n$
such that $\adist_i \in \mtrans(\astate_i)$ and $\adist_i(\astate_{i+1})>0$
for each $0 \leq i < n$.
Let $\last(\afinpath) = \astate_n$ denote the final state of $\afinpath$. 
Let $\mfinpaths{\anmdp}$ be the set of finite paths of the MDP $\anmdp$.  
Let $\mpaths{\anmdp}(\astate)$ and $\mfinpaths{\anmdp}(\astate)$
be the sets of infinite paths and finite paths, respectively,
of $\anmdp$ starting in state $\astate \in \mstates$. 

A \emph{scheduler} is a mapping 
$\asched: \mfinpaths{\anmdp} \ra \dist(\bigcup_{\astate \in \mstates} \mtrans(\astate))$
such that $\asched(\afinpath) \in \dist(\mtrans(\last(\afinpath)))$
for each $\afinpath \in \mfinpaths{\anmdp}$.
Let $\scheds{\anmdp}$ be the set of schedulers of the MDP $\anmdp$.
Given a state $\astate \in \mstates$ and a scheduler $\asched$,
we can define a countably infinite-state DTMC $\adtmcparatwo{\asched}{\astate}$
that corresponds to the behaviour of the scheduler $\asched$ from state $\astate$,
which in turn can be used to define 
a probability measure $\mprob{\asched}{\astate}$ over $\mpaths{\anmdp}(\astate)$
in the standard manner
(see \cite{BK08,FKNP11}).
A scheduler $\asched \in \scheds{\anmdp}$ is \emph{memoryless} if, 
for finite paths $\afinpath, \afinpath' \in \mfinpaths{\anmdp}$ such that $\last(\afinpath) = \last(\afinpath')$,
we have $\asched(\afinpath) = \asched(\afinpath')$.
Let $\memlessscheds{\anmdp}$ be the set of memoryless schedulers of $\anmdp$.
Note that, for a memoryless scheduler $\asched \in \memlessscheds{\anmdp}$,
we can construct a finite DTMC $\afolddtmcpara{\asched} = (\mstates,\folddtrans)$ with
$\folddtrans(\astate,\astate') = \sum_{\adist \in \mtrans(\astate)} \asched(\astate)(\adist) \cdot \adist(\astate')$:
we call this DTMC the \emph{folded DTMC of $\asched$}.
It can be shown that $\mprob{\asched}{\astate}$ and $\dprob{\afolddtmcpara{\asched}}{\astate}$
assign the same probabilities to measurable sets of paths,
because the state $\astate$ of the DTMC $\adtmcparatwo{\asched}{\astate}$
is probabilistic bisimulation equivalent to the state $\astate$ of the folded DTMC $\afolddtmcpara{\asched}$
(for a definition of probabilistic bisimulation and more information on this point, 
see \cite[Section~10.4.2]{BK08}).


\subsubsection{Interval Markov Chains: syntax.}
An (open) \emph{interval Markov chain} (IMC) $\anoimc$ is a pair $(\ostates,\otrans)$,
where $\ostates$ is a finite set of \emph{states},
and $\otrans: \states \times \states \ra \intervals$ 
is a \emph{interval-based transition function}.

In the following, we refer to \emph{edges} as those state pairs for which the transition function
does not assign the probability 0 point interval $[0,0]$.
Formally, let the set of edges $\edges$ of $\anoimc$ be defined as 
$\set{(\astate,\astate') \in \ostates \times \ostates \mid \otrans(\astate,\astate') \neq [0,0]}$.
We use edges to define the notion of path for IMCs:
a path of an IMC $\anoimc$
is a sequence $\astate_0 \astate_1 \cdots$
such that $(\astate_i,\astate_{i+1}) \in \edges$ for all $i \geq 0$.
Given a path $\apath = \astate_0 \astate_1 \cdots$ and $i \geq 0$,
we let $\apath(i) = \astate_i$ be the $(i+1)$-th state along $\apath$.
We use $\gpaths{\anoimc}$ to denote the set of paths of $\anoimc$,
$\gfinpaths{\anoimc}$ to denote the set of finite paths of $\anoimc$,
and $\gpaths{\anoimc}(\astate)$ and $\gfinpaths{\anoimc}(\astate)$
to denote the sets of paths and finite paths starting in state $\astate \in \ostates$.

\sloppypar{
Given a state $\astate \in \ostates$,
we say that a distribution $\apa{\astate} \in \dist(\ostates)$ 
is an \emph{assignment for $\astate$}
if $\apa{\astate}(\astate') \in \otrans(\astate,\astate')$
for each state $\astate' \in \ostates$.
We say that the IMC $\anoimc$ is \emph{well formed} if
there exists at least one assignment for each state.
Note that an assignment for state $\astate \in \ostates$
exists if and only if the following conditions hold:
(1a)~$\sum_{\astate' \in \ostates} \lep{\otrans(\astate,\astate')} \leq 1$,
(1b)~$\sum_{\astate' \in \ostates} \lep{\otrans(\astate,\astate')} = 1$ implies that 
$\otrans(\astate,\astate')$ is left-closed for all $\astate' \in \ostates$,
(2a)~$\sum_{\astate' \in \ostates} \rep{\otrans(\astate,\astate')} \geq 1$,
and 
(2b)~$\sum_{\astate' \in \ostates} \rep{\otrans(\astate,\astate')} = 1$ implies that 
$\otrans(\astate,\astate')$ is right-closed for all $\astate' \in \ostates$.
We henceforth consider IMCs that are well formed.
}
We define the \emph{size of an IMC} $\anoimc = (\ostates,\otrans)$
as the size of the representation of $\otrans$,
which is the sum over all states $\astate,\astate' \in \ostates$
of the binary representation of the endpoints of $\otrans(\astate,\astate')$,
where rational numbers are encoded as the quotient of integers written in binary.


\subsubsection{Interval Markov Chains: semantics.}
%
%
IMCs are typically presented with regard to two semantics, which we consider in turn.
Given an IMC $\anoimc = (\ostates,\otrans)$, 
the \emph{uncertain Markov chain} (UMC) semantics of $\anoimc$, denoted by $\umc{\anoimc}$, 
is the smallest set of DTMCs such that 
$(\ostates,\dtrans) \in \umc{\anoimc}$
if, for each state $\astate \in \ostates$, the distribution $\dtrans(\astate,\cdot)$ is an assignment for $\astate$.
%
%
%
The \emph{interval Markov decision process} (IMDP) semantics of $\anoimc$, denoted by $\imdp{\anoimc}$,
is the MDP $(\ostates,\mtrans)$
where, for each state $\astate \in \ostates$,
we let $\mtrans(\astate)$ 
be the set of assignments for $\astate$.


\subsubsection{Reachability.}
Let $\anoimc = (\ostates,\otrans)$ be an IMC
and let $\targetset \subseteq \ostates$ be a set of states.
We define $\reachpaths{\targetset} \subseteq \gpaths{\anoimc}$ to be the set of paths of $\anoimc$
that reach at least one state in $\targetset$.
Formally, $\reachpaths{\targetset} = 
\set{ \apath \in \gpaths{\anoimc} \mid \exists i \in \Nset . \apath(i) \in \targetset }$.
In the following we assume without loss of generality that states in $\targetset$ are absorbing
in all the IMCs that we consider,
i.e., $\otrans(\astate,\astate)=[1,1]$ for all states $\astate \in \targetset$.


\subsubsection{Edge sets.}
Let $\anoimc = (\ostates,\otrans)$ be an IMC.
Let $\astate \in \ostates$ be a state of $\anoimc$,
and let $\edgesfrom{\astate} = \set{ (\astate,\astate') \in \edges \mid  \astate' \in \ostates}$
be the set of edges of $\anoimc$ with source $\astate$.
Let $\star \in \set{ \lcrc, \lorc, \lcro, \loro, \lnz, \lcz, \loz, \lz }$,
and let 
$\edgespara{\star} = \set{(\astate,\astate') \in \edges \mid \otrans(\astate,\astate') \in \intervals^\star}$.
Given $\astateset \subseteq \ostates$, and given $\astate$ and $\star$ as defined above,
let $\edgesfromtwo{\astate}{\astateset} =  
\set{ (\astate,\astate') \in \edgesfrom{\astate} \mid  \astate' \in \astateset}$
and $\edgesfromthree{\star}{\astate}{\astateset} = 
\edgesfromtwo{\astate}{\astateset} \cap \edgespara{\star}$.


\subsubsection{Valid edge sets.}
We are interested in identifying the sets of edges from state $\astate \in \ostates$
that result from assignments.
Such a set is characterised by two syntactic conditions:
the first condition requires that the sum of the upper bounds of the set's edges' intervals is at least 1,
whereas the second condition specifies the edges from state $\astate$ 
that are \emph{not} included in the set can be assigned probability 0.
Formally, we say that a non-empty subset $\edgesubset \subseteq \edgesfrom{\astate}$
of edges from $\astate$ is
\emph{large} if either 
(a)~$\sum_{\anedge \in \edgesubset} \rep{\otrans(\anedge)} > 1$
or (b)~$\sum_{\anedge \in \edgesubset} \rep{\otrans(\anedge)} = 1$ and 
$\edgesubset \subseteq \edgespara{\rc}$.
The set $\edgesubset$ is \emph{realisable} 
if $\edgesfrom{\astate} \setminus \edgesubset \subseteq \edgespara{\lcz}$.
Then we say that $\edgesubset \subseteq \edgesfrom{\astate}$ is \emph{valid} 
if it is large and realisable.
The following lemma specifies that a valid edge set for state $\astate$ 
characterises exactly the support sets of some assignments for $\astate$.

\begin{lemma}\label{lem:validass}
Let $\astate \in \ostates$ and $\edgesubset \subseteq \edgesfrom{\astate}$.
Then $\edgesubset$ is valid if and only if
there exists an assignment $\apa{\astate}$ for $\astate$
such that $\set{(\astate,\astate') \mid \astate' \in \support(\apa{\astate})} = \edgesubset$.
\end{lemma}

A consequence of \lemref{lem:validass} is that, 
because we consider only well-formed IMCs,
there exists at least one valid subset of outgoing edges from each state.

For each state $\astate \in \ostates$,
we let $\validsetsfrom{\astate} = \set{ \edgesubset \subseteq \edgesfrom{\astate} \mid \edgesubset \mbox{ is valid}}$.
Note that, in the worst case (when all edges in $\edgesfrom{\astate}$ are asociated with intervals $[0,1]$), 
$|\validsetsfrom{\astate}| = 2^{|\edgesfrom{\astate}|}-1$.
Let $\validsets = \bigcup_{\astate \in \ostates} \validsetsfrom{\astate}$
be the set of valid sets of the IMC.
Given a valid set $\edgesubset \in \validsets$,
we let $\possass{\edgesubset}$ be the set of assignments $\apa{\astate}$ 
that witness \lemref{lem:validass},
i.e., all assignments $\apa{\astate}$ 
such that $\set{(\astate,\astate') \mid \astate' \in \support(\apa{\astate})} = \edgesubset$.
A \emph{witness assignment function} $\waf: \validsets \ra \dist(\ostates)$
assigns to each valid set $\edgesubset \in \validsets$ an assignment from $\possass{\edgesubset}$.

\begin{example}
For the state $\astate_1$ of the IMC $\anoimc_2$ of \figref{fig:general_example},
the valid edge sets are 
$\edgesubset_1 = \set{(\astate_1,\astate_0),(\astate_1,\astate_1),(\astate_1,\astate_2)}$
and $\edgesubset_2 = \set{(\astate_1,\astate_0),(\astate_1,\astate_2)}$,
reflecting the intuition that the edge $(\astate_1,\astate_1)$ can be assigned (exactly) probability $0$.
Note that reducing the right endpoint of $(\astate_1,\astate_0)$ to $0.7$
would result in $\edgesubset_1$
being the only valid set associated with $\astate_1$,
because $\edgesubset_2$ would not be large.
An example of a witness assignment function $\waf$ for state $\astate_1$ of $\anoimc_2$ 
is $\waf(\edgesubset_1)(\astate_0) = 0.7$, $\waf(\edgesubset_1)(\astate_1) = 0.12$
and $\waf(\edgesubset_1)(\astate_2) = 0.18$,
and $\waf(\edgesubset_2)(\astate_0) = 0.8$ and $\waf(\edgesubset_2)(\astate_2) = 0.2$.
\end{example}


\subsubsection{Qualitative MDP abstractions.}
The \emph{qualitative MDP abstraction of $\anoimc$ with respect to witness assignment function $\waf$}
is the MDP $\qa{\anoimc}{\waf} = (\ostates,\qatrans{\waf})$,
where $\qatrans{\waf}$ is defined by 
$\qatrans{\waf}(\astate) = \set{ \waf(\edgesubset) \mid \edgesubset \in \validsetsfrom{\astate}}$
for each state $\astate \in \ostates$.

\section{Qualitative Reachability: UMC semantics}\label{sec:umc_algos}

Qualitative reachability problems can be classified into four categories,
depending on whether the probability of reaching the target set $\targetset$ is $0$ or $1$
for some or for all ways of assigning probabilities to intervals.
For the UMC semantics, we consider the computation of the following sets:
\begin{itemize}
\item
$\aqset{0}{\uident} =
\set{ \astate \in \ostates \mid 
\forall \adtmc \in \umc{\anoimc} \ldot \dprob{\adtmc}{\astate}(\reachpaths{\targetset}) = 0 }$;
\item
$\eqset{0}{\uident} = 
\set{ \astate \in \ostates \mid 
\exists \adtmc \in \umc{\anoimc} \ldot \dprob{\adtmc}{\astate}(\reachpaths{\targetset}) = 0 }$;
\item
$\eqset{1}{\uident} = 
\set{ \astate \in \ostates \mid 
\exists \adtmc \in \umc{\anoimc} \ldot \dprob{\adtmc}{\astate}(\reachpaths{\targetset}) = 1 }$;
\item
$\aqset{1}{\uident} = 
\set{ \astate \in \ostates \mid 
\forall \adtmc \in \umc{\anoimc} \ldot \dprob{\adtmc}{\astate}(\reachpaths{\targetset}) = 1 }$.
\end{itemize}

The remainder of this section is dedicated to showing the following result.

\begin{theorem}\label{thm:umcs}
The sets $\aqset{0}{\uident}$, $\eqset{0}{\uident}$, 
$\eqset{1}{\uident}$ and $\aqset{1}{\uident}$ can be computed 
in polynomial time in the size of the IMC.
\end{theorem}

\subsubsection{Computation of $\aqset{0}{\uident}$.}
The case for $\aqset{0}{\uident}$ is straightforward.
We compute the state set
$\ostates \setminus \aqset{0}{\uident} = 
\set{ \astate \in \ostates \mid 
\exists \adtmc \in \umc{\anoimc} \ldot \dprob{\adtmc}{\astate}(\reachpaths{\targetset}) > 0 }$,
which reduces to reachability on the graph of the IMC according to the following lemma.

\begin{lemma}\label{lem:reach_umc}
Let $\astate \in \ostates$.
There exists $\adtmc \in \umc{\anoimc}$ such that $\dprob{\adtmc}{\astate}(\reachpaths{\targetset}) > 0$
if and only if there exists a path $\afinpath \in \gfinpaths{\anoimc}(\astate)$ such that $\last(\afinpath) \in \targetset$. 
\end{lemma}
Hence the set $\aqset{0}{\uident}$
is equal to the complement of the set of states from which there exists a path reaching $\targetset$
in the graph of the IMC (that is, the graph $(\ostates,\edges)$).
Given that the latter set of states can be computed in polynomial time,
we conclude that $\aqset{0}{\uident}$ can be computed in polynomial time.

\subsubsection{Computation of $\eqset{0}{\uident}$.}
We show that $\eqset{0}{\uident}$
can be obtained by computing the set of states 
from which there exists a scheduler for which $\targetset$ is reached with probability 0
in the qualitative MDP abstraction $\qa{\anoimc}{\waf} = (\ostates,\qatrans{\waf})$
of $\anoimc$ with respect to some (arbitrary) witness asignment function $\waf$.

First we establish that the set of states of $\qa{\anoimc}{\waf}$ 
for which there exists a scheduler such that $\targetset$ is reached with probability 0 (respectively, probability 1) 
is equal to the set of states of $\anoimc$ for which there exists a DTMC in $\umc{\anoimc}$
such that $\targetset$ is reached with probability 0 (respectively, probability 1).

\begin{lemma}\label{lem:qa-umc}
Let $\astate \in \ostates$, $\bowtie \in \set{<, =, >}$ and $\lambda  \in \set{0,1}$.
There exists $\adtmc \in \umc{\anoimc}$ 
such that $\dprob{\adtmc}{\astate}(\reachpaths{\targetset}) \bowtie \lambda$
if and only if there exists a scheduler $\asched \in \scheds{\qa{\anoimc}{\waf}}$ 
such that $\mprob{\asched}{\astate}(\reachpaths{\targetset}) \bowtie \lambda$. 
\end{lemma}

In particular, \lemref{lem:qa-umc} allows us to reduce the problem of computing $\eqset{0}{\uident}$
to that of computing the set 
$\set{ \astate \in \ostates \mid \exists \asched \in \scheds{\qa{\anoimc}{\waf}}
\ldot
\mprob{\asched}{\astate}(\reachpaths{\targetset}) = 0 }$ on $\qa{\anoimc}{\waf}$.
As in the case of standard finite MDP techniques (see \cite{FKNP11}),
we proceed by computing the \emph{complement} of this set,
i.e., we compute the set
$\set{ \astate \in \ostates \mid \forall \asched \in \scheds{\qa{\anoimc}{\waf}}
\ldot
\mprob{\asched}{\astate}(\reachpaths{\targetset}) > 0 }$.
For a set $\astateset \subseteq \ostates$,
let $\cpre(\astateset) = 
\set{ \astate \in \ostates \mid  
\exists \adist \in \qatrans{\waf}(\astate) \ldot \support(\adist) \subseteq \astateset  }$
be the set of states for which there exists a distribution such that 
all states assigned positive probability by the distribution are in $\astateset$.
Furthermore, we let 
$\dualcpre(\astateset) =
\set{ \astate \in \ostates \mid  
\forall \adist \in \qatrans{\waf}(\astate) \ldot \support(\adist) \cap \astateset \neq \emptyset  }$
be the dual of the $\cpre$ operator 
(i.e., $\cpre(\astateset) = \ostates \setminus \dualcpre(\ostates \setminus \astateset)$),
that is the set of states from which it is inevitable to make a transition to $\astateset$ with positive probability.
The standard algorithm for computing the set of states of a finite MDP for which
all schedulers are such that a set $\targetset$ of target states 
is reached with probability strictly greater than 0 
operates in the following way:
starting from $\astateset_0 = \targetset$,
we let $\astateset_{i+1} = \astateset_i \cup \dualcpre(\astateset_i)$
for progressively larger values of $i \geq 0$,
until we reach a fixpoint 
(that is, until we obtain $\astateset_{i^*+1} = \astateset_{i^*}$ for some $i^*$).
However, a direct application of this algorithm to $\qa{\anoimc}{\waf}$
would result in an exponential-time algorithm, 
given that the size of the transition function $\qatrans{\waf}$ of $\qa{\anoimc}{\waf}$
may be exponential in the size of $\anoimc$.
For this reason, we propose an algorithm that operates directly on the IMC $\anoimc$,
without needing the explicit construction of $\qa{\anoimc}{\waf}$.
We proceed by establishing that 
$\cpre$ can be implemented in polynomial time in the size of $\anoimc$.

\begin{lemma}\label{lem:cpre_eff}
Let $\astate \in \ostates$ and $\astateset \subseteq \ostates$.
Then $\astate \in \cpre(\astateset)$ if and only if
(1)~$\edgesfromtwo{\astate}{\ostates \setminus \astateset} \subseteq \edgespara{\lcz}$, and
(2)~$\edgesfromtwo{\astate}{\astateset}$ is large.
The set $\cpre(\astateset)$ can be computed in polynomial time in the size of the IMC $\anoimc$.
\end{lemma}
The intuition underlying \lemref{lem:cpre_eff} is that 
conditions~(1) and (2) encode realisibility and largeness, i.e., validity,
of edge set $\edgesfromtwo{\astate}{\astateset}$. 
From \lemref{lem:validass},
their satisfaction means that there exists a distribution in $\qatrans{\waf}(\astate)$
with support set equal to the set of target states of edges in $\edgesfromtwo{\astate}{\astateset}$.
We consider the largest edge set with target states in $\astateset$,
i.e., $\edgesfromtwo{\astate}{\astateset}$,
because taking smaller edge sets with targets in $\astateset$
would make the conditions~(1) and (2) more difficult to satisfy.

The final part of \lemref{lem:cpre_eff} follows from the fact that 
conditions~(1) and (2) in \lemref{lem:cpre_eff} can be checked 
in polynomial time in the size of $\anoimc$.
Hence our algorithm avoids the construction of the qualitative MDP abstraction $\qa{\anoimc}{\waf}$,
and instead consists of direct computation of the sets 
$\astateset_0 = \targetset$ and 
$\astateset_{i+1} = \astateset_i \cup \ostates \setminus \cpre(\ostates \setminus \astateset_i)$
for increasing indices $i$ until a fixpoint is reached.
Given that a fixpoint must be reached within $|\ostates|$ steps,
and the computation of $\cpre(\astateset_i)$ can be done in polynomial time in the size of $\anoimc$,
we have that the set 
$\set{ \astate \in \ostates \mid \forall \asched \in \scheds{\qa{\anoimc}{\waf}}
\ldot
\mprob{\asched}{\astate}(\reachpaths{\targetset}) > 0 }$
can be computed 
in polynomial time in the size of $\anoimc$.
The complement of this set is equal to $\eqset{0}{\uident}$,
as established by \lemref{lem:qa-umc},
and hence we can compute $\eqset{0}{\uident}$ 
in polynomial time in the size of $\anoimc$.

\subsubsection{Computation of $\eqset{1}{\uident}$.}
We proceed in a manner analogous to that for the case of $\eqset{0}{\uident}$.
First we note that, by \lemref{lem:qa-umc},
we have that $\eqset{1}{\uident}$
is equal to the set of states of $\qa{\anoimc}{\waf}$ such that 
there exists a scheduler for which $\targetset$ is reached with probability $1$.
Hence, our aim is to compute the set 
$\set{ \astate \in \ostates \mid \exists \asched \in \scheds{\qa{\anoimc}{\waf}}
\ldot
\mprob{\asched}{\astate}(\reachpaths{\targetset}) = 1 }$
on $\qa{\anoimc}{\waf}$.
We recall the standard algorithm for the computation of this set on finite MDPs \cite{deAlf97,deAlf99}.
Given state sets $\astateset,\anotherstateset \subseteq \ostates$,
we let
\[
\apre(\anotherstateset,\astateset) = 
\set{ \astate \in \ostates \mid
\exists \adist \in \qatrans{\waf}(\astate) \ldot 
\support(\adist) \subseteq \anotherstateset \wedge \support(\adist) \cap \astateset \neq \emptyset }
\]
be the set of states for which there exists a distribution such that 
all states assigned positive probability by the distribution are in $\anotherstateset$
and there exists a state assigned positive probability by the distribution that is in $\astateset$.
The standard algorithm proceeds by setting $\anotherstateset_0 = \ostates$ and $\astateset_0^0 = \targetset$.
Then the sequence $\astateset_0^0, \astateset_1^0, \cdots$
is computed by letting 
$\astateset_{i_0+1}^0 = \astateset_{i_0}^0 \cup \apre(\anotherstateset_0,\astateset_{i_0}^0)$
for progressively larger indices $i_0 \geq 0$ until a fixpoint is obtained,
that is, until we obtain $\astateset_{i^*_0+1}^0 = \astateset_{i^*_0}^0$ for some $i^*_0$.
Next we let $\anotherstateset_1 = \astateset_{i_0^*}^0$, $\astateset_0^1 = \targetset$
and compute $\astateset_{i_1+1}^1 = \astateset_{i_1}^1 \cup \apre(\anotherstateset_1,\astateset_{i_1}^1)$
for larger $i_1 \geq 0$ until a fixpoint $ \astateset_{i^*_1}^1$ is obtained.
Then we let $\anotherstateset_2 = \astateset_{i_1^*}^1$ and $\astateset_0^2 = \targetset$,
and repeat the process.
We terminate the algorithm when a fixpoint is reached in the sequence 
$\anotherstateset_0, \anotherstateset_1, \cdots$.\footnote{
Readers familiar with $\mu$-calculus will observe that the algorithm 
can be expressed using the term 
$\nu \anotherstateset \ldot \mu \astateset (\targetset \cup  \apre(\anotherstateset,\astateset))$
\cite{deAlf99}.
}
The algorithm requires at most $|\ostates|^2$ calls to $\apre$.
In an analogous manner to $\cpre$ in the case of $\eqset{0}{\uident}$,
we show that $\apre$ can characterised by efficiently checkable conditions on $\anoimc$.

\begin{lemma}\label{lem:apre_eff}
Let $\astate \in \ostates$ and let $\astateset, \anotherstateset \subseteq \ostates$.
Then $\astate \in \apre(\anotherstateset, \astateset)$ if and only if
(1)~$\edgesfromtwo{\astate}{\astateset \cap \anotherstateset} \ne \emptyset$, 
(2)~$\edgesfromtwo{\astate}{\ostates \setminus \anotherstateset} \subseteq \edgespara{\lcz}$,
and 
(3)~$\edgesfromtwo{\astate}{\anotherstateset}$ is large.
The set $\apre(\anotherstateset,\astateset)$ can be computed 
in polynomial time in the size of the IMC $\anoimc$.
\end{lemma}
The intution underlying \lemref{lem:apre_eff} is similar to that of \lemref{lem:cpre_eff}.

\sloppypar{
Hence we obtain an overall polynomial-time algorithm
for computing $\set{ \astate \in \ostates \mid \exists \asched \in \scheds{\qa{\anoimc}{\waf}}
\ldot
\mprob{\asched}{\astate}(\reachpaths{\targetset}) = 1 }$
which, from \lemref{lem:qa-umc}, equals $\eqset{1}{\uident}$.
}

\subsubsection{$\aqset{1}{\uident}$.}
We recall the standard algorithm for determining the set of states
for which all schedulers reach a target set with probability $1$ on a finite MDP (see \cite{FKNP11}):
from the set of states of the MDP, 
we first remove states from which the target state can be reached with probability $0$ 
(for some scheduler),
then successively remove states for which it is possible to reach a previously removed state with positive probability.
For each of the remaining states, there exists a scheduler that can reach the target set with probability $1$.

We propose an algorithm for IMCs that is inspired by this standard algorithm for finite MDPs.
Our aim is to compute the complement of $\aqset{1}{\uident}$, i.e., the state set
$\ostates \setminus \aqset{1}{\uident}
=
\set{ \astate \in \ostates \mid 
\exists \adtmc \in \umc{\anoimc} \ldot \dprob{\adtmc}{\astate}(\reachpaths{\targetset}) < 1 }$.

\begin{lemma}\label{lem:min_one_umc}
Let $\astate \in \ostates$.
There exists $\adtmc \in \umc{\anoimc}$ such that $\dprob{\adtmc}{\astate}(\reachpaths{\targetset}) < 1$
if and only if there exists a path $\afinpath \in \gfinpaths{\anoimc}(\astate)$ such that $\last(\afinpath) \in \eqset{0}{\uident}$. 
\end{lemma}
Hence the set $\aqset{1}{\uident}$ can be computed by taking the complement 
of the set of states for which there exists a path to $\eqset{0}{\uident}$ in the graph of $\anoimc$.
Given that $\eqset{0}{\uident}$, and the set of states reaching $\eqset{0}{\uident}$, 
can be computed in polynomial time,
we have obtained a polynomial-time algorithm for computing $\aqset{1}{\uident}$.
Together with the cases for $\aqset{0}{\uident}$, $\eqset{0}{\uident}$ and $\eqset{1}{\uident}$,
this establishes \thmref{thm:umcs}.

\section{Qualitative Reachability: IMDP semantics}\label{sec:imdp_algos}

We now focus on the IMDP semantics,
and consider the computation of the following sets:
\begin{itemize}
\item
$\aqset{0}{\iident} = 
\set{ \astate \in \ostates \mid 
\forall \asched \in \scheds{\imdp{\anoimc}} \ldot \mprob{\asched}{\astate}(\reachpaths{\targetset}) = 0 }$;
\item
$\eqset{0}{\iident} = 
\set{ \astate \in \ostates \mid 
\exists \asched \in \scheds{\imdp{\anoimc}} \ldot \mprob{\asched}{\astate}(\reachpaths{\targetset}) = 0 }$;
\item
$\eqset{1}{\iident} = 
\set{ \astate \in \ostates \mid 
\exists \asched \in \scheds{\imdp{\anoimc}} \ldot \mprob{\asched}{\astate}(\reachpaths{\targetset}) = 1 }$;
\item
$\aqset{1}{\iident} = 
\set{ \astate \in \ostates \mid 
\forall \asched \in \scheds{\imdp{\anoimc}} \ldot \mprob{\asched}{\astate}(\reachpaths{\targetset}) = 1 }$.
\end{itemize}

This section will be dedicated to showing the following result.
We note that the cases for $\aqset{0}{\iident}$, $\eqset{0}{\iident}$ and $\eqset{1}{\iident}$
proceed in a manner similar to the UMC case 
(using either graph reachability or reasoning based on the qualitative MDP abstraction);
instead the case for $\aqset{1}{\iident}$ requires substantially different techniques.

\begin{theorem}
The sets $\aqset{0}{\iident}$, $\eqset{0}{\iident}$, 
$\eqset{1}{\iident}$ and $\aqset{1}{\iident}$ can be computed 
in polynomial time in the size of the IMC.
\end{theorem}

\subsubsection{Computation of $\aqset{0}{\iident}$.}
As in the case of UMCs,
the computation of $\aqset{0}{\iident}$ reduces to straightforward reachability analysis on the graph of the IMC $\anoimc$.
The correctness of the reduction is based on the following lemma.

\begin{lemma}\label{lem:reach_imc}
Let $\astate \in \ostates$.
There exists $\asched \in \scheds{\imdp{\anoimc}}$ 
such that $\mprob{\asched}{\astate}(\reachpaths{\targetset}) > 0$
if and only if there exists a path $\afinpath \in \gfinpaths{\anoimc}(\astate)$ 
such that $\last(\afinpath) \in \targetset$. 
\end{lemma}

Therefore, to obtain $\aqset{0}{\iident}$,
we proceed by computing the state set
$\ostates \setminus \aqset{0}{\iident} = 
\set{ \astate \in \ostates \mid 
\exists \asched \in \scheds{\imdp{\anoimc}} \ldot \mprob{\asched}{\astate}(\reachpaths{\targetset}) > 0 }$,
which reduces to reachability on the graph of the IMC according to \lemref{lem:reach_imc},
and then taking the complement.

\subsubsection{Computation of $\eqset{0}{\iident}$ and $\eqset{1}{\iident}$.}
In the following we fix an arbitrary witness assignment function $\waf$ of $\anoimc$.
\lemref{lem:exists_imdp}
establishes that $\eqset{0}{\iident}$ (respectively, $\eqset{1}{\iident}$)
equals the set of states of the qualitative MDP abstraction $\qa{\anoimc}{\waf}$ 
with respect to $\waf$
for which there exists some scheduler such that $\targetset$ is reached with probability $0$
(respectively, probability $1$).

\begin{lemma}\label{lem:exists_imdp}
Let $\astate \in \ostates$ and $\lambda \in \set{0,1}$.
There exists $\asched \in \scheds{\imdp{\anoimc}}$ 
such that $\mprob{\asched}{\astate}(\reachpaths{\targetset}) = \lambda$
if and only if there exists a scheduler $\asched' \in \scheds{\qa{\anoimc}{\waf}}$ 
such that $\mprob{\asched'}{\astate}(\reachpaths{\targetset}) = \lambda$. 
\end{lemma}

Given that we have shown in Section~\ref{sec:umc_algos} that the 
set of states of the qualitative MDP abstraction $\qa{\anoimc}{\waf}$
for which there exists some scheduler such that $\targetset$ is reached with probability $0$
(respectively, probability $1$)
can be computed in polynomial time in the size of $\anoimc$,
we obtain polynomial-time algorithms for computing $\eqset{0}{\iident}$ 
(respectively, $\eqset{1}{\iident}$).


\subsubsection{Computation of $\aqset{1}{\iident}$.}
This case is notably different from the other three cases for the IMDP semantics,
because schedulers that are \emph{not} memoryless 
may influence whether a state is included in $\aqset{1}{\iident}$.
In particular, we recall the example of the IMC of Figure \ref{fig:open_example}:
as explained in Section \ref{sec:intro},
we have $\astate_0 \not\in \aqset{1}{\iident}$.
In contrast, we have $\astate_0 \in \aqset{1}{\uident}$,
and $\astate_0$ would be in $\aqset{1}{\iident}$ if we restricted the IMDP semantics to memoryless
(actually finite-memory, in this case) schedulers.
For this reason, a qualitative MDP abstraction is not useful for computing $\aqset{1}{\iident}$,
because it is based on the use of witness assignment functions that assign \emph{constant} probabilities to 
sets of edges available from states:
on repeated visits to a state, the (finite) set of available distributions remains the same
in a qualitative MDP abstraction.
Therefore we require alternative analysis methods that are not based on the qualitative MDP abstraction.
Our approach is based on the notion of end components,
which is a standard concept in the field of MDP verification \cite{deAlf97}.
In this section we introduce an alternative notion of end components,
defined solely in terms of states of the IMC, 
which characterises situations in which the IMC can confine its behaviour to certain state sets
with positive probability in the IMDP semantics
(for example, the IMC of Figure \ref{fig:open_example} can confine itself to state $\astate_0$
with positive probability in the IMDP semantics).

An \emph{IMC-level end component} (ILEC) is a set $\ilecstates \subseteq \ostates$ of states
that is strongly connected and
such that the total probability assigned to edges that have a source state in $\ilecstates$ 
but a target state outside of $\ilecstates$
can be made to be arbitrarily small
(note that such edges must have an interval with a left endpoint of $0$).
Formally, $\ilecstates \subseteq \ostates$ is an ILEC if, for each state $\astate \in \ilecstates$,
we have (1)~$\edgesfromthree{\lnz}{\astate}{\ostates \setminus \ilecstates} = \emptyset$,
(2)~$\sum_{\anedge \in \edgesfromtwo{\astate}{\ilecstates}} \rep{\otrans(\anedge)} \geq 1$,
and (3)~the graph $(\ilecstates,\edges \cap (\ilecstates \times \ilecstates))$ is strongly connected.

\begin{example}
In the IMC $\anoimc_1$ of \figref{fig:open_example}, 
the set $\set{\astate_0}$ is an ILEC:
for condition~(1), the edge $(\astate_0,\astate_1)$ 
(the only edge in $\edgesfromtwo{\astate_0}{\ostates \setminus \set{\astate_0}}$)
is not in $\edgespara{\lnz}$, and,
for condition~(2), we have $\rep{\otrans(\astate_0,\astate_1)} = 1$.
In the IMC $\anoimc_2$ of \figref{fig:general_example}, 
the set $\set{\astate_0,\astate_1}$ is an ILEC:
for condition~(1), the only edge leaving $\set{\astate_0,\astate_1}$ has $0$ as its left endpoint,
i.e., $\otrans(\astate_1,\astate_2)=(0,0.2]$, hence 
$\edgesfromthree{\lnz}{\astate_0}{\set{\astate_2}} = 
\edgesfromthree{\lnz}{\astate_1}{\set{\astate_2}} = \emptyset$;
for condition~(2), we have 
$\rep{\otrans(\astate_0,\astate_0)}+\rep{\otrans(\astate_0,\astate_1)} = 1.6 \geq 1$
and 
$\rep{\otrans(\astate_1,\astate_0)}+\rep{\otrans(\astate_1,\astate_1)} = 1.3 \geq 1$.
In both cases, the identified sets clearly induce strongly connected subgraphs,
thus satisfying condition~(3).
\end{example}

\begin{remark}
Both conditions (1) and (2) are necessary to ensure that the probability of 
leaving $\ilecstates$ in one step can be made arbitrarily small.
Consider an IMC with state $\astate \in \ilecstates$ such that 
$\edgesfromtwo{\astate}{\ilecstates} = \set{ \anedge_1 }$ and
$\edgesfromtwo{\astate}{\ostates \setminus \ilecstates} = \set{ \anedge_2, \anedge_3}$,
where $\otrans(\anedge_1)=[0.6,0.8]$, 
$\otrans(\anedge_2)=[0,0.2]$ and $\otrans(\anedge_3)=[0,0.2]$.
Then condition (1) holds but condition (2) does not:
indeed, at least total probability $0.2$ must be assigned to the edges ($\anedge_2$ and $\anedge_3$)
that leave $\ilecstates$.
Now consider an IMC with state $\astate \in \ilecstates$ such that 
$\edgesfromtwo{\astate}{\ilecstates} = \set{ \anedge_1, \anedge_2 }$ and
$\edgesfromtwo{\astate}{\ostates \setminus \ilecstates} = \set{ \anedge_3}$,
where $\otrans(\anedge_1)=[0,0.5]$, $\otrans(\anedge_2)=[0,0.5]$ 
and $\otrans(\anedge_3)=[0.1,0.5]$.
Then condition (2) holds (because the sum of the right endpoints of the intervals
associated with $\anedge_1$ and $\anedge_2$ is equal to $1$),
but condition (1) does not
(because the interval associated with $\anedge_3$ specifies that probability at least $0.1$
must be assigned to leaving $\ilecstates$).
Note also that if $\edgesfromtwo{\astate}{\ilecstates} \subseteq \edgespara{\lcro} \cup \edgespara{\loro}$
(all edges in $\edgesfromtwo{\astate}{\ilecstates}$ have right-open intervals)
and $\sum_{\anedge \in \edgesfromtwo{\astate}{\ilecstates}} \rep{\otrans(\anedge)} = 1$,
there must exist a least one edge in $\edgesfromtwo{\astate}{\ostates \setminus \ilecstates}$
by well formedness.
\end{remark}

Let $\ilecs$ be the set of ILECs of $\anoimc$.
We say that an ILEC $\ilecstates \in \ilecs$ is \emph{maximal} 
if there does not exist any $\ilecstates' \in \ilecs$ such that $\ilecstates \subset \ilecstates'$.
For a path $\apath \in \mpaths{\imdp{\anoimc}}(\astate)$,
let $\infstatesalong{\apath} \subseteq \ostates$ be the states
that appear infinitely often along $\apath$,
i.e., for $\apath = \astate_0 \adist_0 \astate_1 \adist_1 \cdots$,
we have $\infstatesalong{\apath} = 
\set{ \astate \in \ostates \mid \forall i \in \Nset \ldot \exists j>i \ldot \astate_j = \astate }$.
We present a result for ILECs that is analogous to 
the fundamental theorem of end components of \cite{deAlf97}:
the result specifies that, with probability $1$,
a scheduler of the IMDP semantics of $\anoimc$ must confine itself to an ILEC.

\begin{lemma}\label{lem:fund_ilecs}
For $\astate \in \ostates$ and $\asched \in \scheds{\imdp{\anoimc}}$,
we have $\mprob{\asched}{\astate}(\set{ \apath \mid \infstatesalong{\apath} \in \ilecs}) = 1$.
\end{lemma}

We now show that there exists a scheduler that, from a state within an ILEC,
can confine the IMC to the ILEC with positive probability.
This result is the ILEC analogue of a standard result for end components of finite MDPs
that specifies that there exists a scheduler that, from a state of an end component,
can confine the MDP to the end component with probability $1$
(see \cite{deAlf97,BK08}).
In the case of IMCs and ILECs, 
it is not possible to obtain an analogous result for probability $1$;
in the example of Figure \ref{fig:open_example},
the singleton set $\set{ \astate_0 }$ is an ILEC, 
but it is not possible to find a scheduler that remains in $\astate_0$ with probability $1$,
because with each transition the IMC goes to $\astate_1$ with positive probability.
For our purposes, it is sufficient to have a result stating that, from an ILEC,
the IMC can be confined to the ILEC with positive probability.

\begin{lemma}\label{lem:remain_ilec}
\sloppypar{Let $\ilecstates \in \ilecs$ and $\astate \in \ilecstates$.
There exists $\asched \in \scheds{\imdp{\anoimc}}$
such that $\mprob{\asched}{\astate}(\set{ \apath \mid 
\apath \not\in \reachpaths{\ostates \setminus \ilecstates} \wedge \infstatesalong{\apath} = \ilecstates}) > 0$.}
\end{lemma}
The key point of the proof of \lemref{lem:remain_ilec}
is the definition of a scheduler 
that assigns progressively decreasing probability to all edges in $\edgespara{\lz}$ that leave ILEC $\ilecstates$,
in such a way as to guarantee that the IMC is confined in $\ilecstates$ with positive probability.
This is possible because condition~(2) of the definition of ILECs
specifies that there is no lower bound on the probability that must 
be assigned to edges that leave $\ilecstates$.
Furthermore, the scheduler is defined so that 
the remaining probability at each step that is assigned between all edges that stay in $\ilecstates$
is always no lower than some fixed lower bound;
this characteristic of the scheduler,
combined with the fact that we remain in $\ilecstates$ with positive probability
and the fact that $\ilecstates$ is strongly connected,
means that we visit all states of $\ilecstates$ with positive probability under the defined scheduler.

Let $\statesettwo{\ilecs}{\targetset} = 
\bigcup \set{ \ilecstates \in \ilecs \mid \ilecstates \cap \targetset = \emptyset }$
be the union of states of ILECs that do not contain states in $\targetset$.
Using \lemref{lem:fund_ilecs} and \lemref{lem:remain_ilec} in a standard way,
we can show that the existence of a scheduler of $\imdp{\anoimc}$ 
that reaches $\targetset$ with probability strictly less than $1$
is equivalent to the existence of a path in the graph of $\anoimc$ 
that reaches $\statesettwo{\ilecs}{\targetset}$.

\begin{proposition}\label{prop:imdp_ltone}
Let $\astate \in \ostates$.
There exists $\asched \in \scheds{\imdp{\anoimc}}$ 
such that $\mprob{\asched}{\astate}(\reachpaths{\targetset}) < 1$
if and only if there exists a finite path $\afinpath \in \gfinpaths{\anoimc}(\astate)$ 
such that $\last(\afinpath) \in \statesettwo{\ilecs}{\targetset}$.
\end{proposition}

Hence we identify the set $\aqset{1}{\iident}$
by computing the complement of $\aqset{1}{\iident}$,
i.e., the set
$\ostates \setminus \aqset{1}{\iident}
= 
\set{ \astate \in \ostates \mid 
\forall \asched \in \scheds{\imdp{\anoimc}} \ldot \mprob{\asched}{\astate}(\reachpaths{\targetset}) < 1 }$.
Using \propref{prop:imdp_ltone},
this set can be computed by considering reachability 
on the graph of $\anoimc$ of the set $\statesettwo{\ilecs}{\targetset}$.
The set $\statesettwo{\ilecs}{\targetset}$ can be computed in polynomial time in the size of $\anoimc$
in a manner similar to the computation of maximal end components of MDPs (see \cite{deAlf97,BK08}).
First we compute all strongly connected components 
$(\ilecstates_1,\edges \cap (\ilecstates_1 \times \ilecstates_1)), \cdots, 
(\ilecstates_m,\edges \cap (\ilecstates_m \times \ilecstates_m))$ 
of the graph 
$(\ostates \setminus \targetset,
\edges \cap ((\ostates \setminus \targetset) \times (\ostates \setminus \targetset)))$ of $\anoimc$.
Then, for each $1 \leq i \leq m$, we remove from $\ilecstates_i$ all states for which
conditions (1) or (2) in the definition of ILECs do \emph{not} hold with respect to $\ilecstates_i$
(these conditions can be checked in polynomial time for each state),
to obtain the state set $\ilecstates_i'$.
Next, we compute the strongly connected components of the graph 
$(\ilecstates_i',\edges  \cap (\ilecstates_i' \times \ilecstates_i'))$,
and for each of these, repeat the procedure described above.
We terminate the algorithm when it is not possible to remove a state
(via a faliure to satisfy a least one of the conditions (1) and (2) in the definition of ILECs)  
from any generated strongly connected component.
The generated state sets of the strongly connected components obtained 
will be be the maximal ILECs that do not contain any state in $\targetset$,
and their union is $\statesettwo{\ilecs}{\targetset}$.
Hence the overall algorithm for computing $\aqset{1}{\iident}$
is in polynomial time in the size of $\anoimc$.

\section{Conclusion}

We have presented algorithms for qualitative reachability properties for open IMCs.
In the context of qualitative properties of system models with \emph{fixed} probabilities
on their transitions,
probability can be regarded as imposing a fairness constraint,
i.e., paths for which a state is visited infinitely often and one of its successors is visited only finitely often
have probability $0$.
In open IMCs, the possibility to make the probability of a transition converge to $0$ in the IMDP semantics
captures a different phenomenon, 
which is key for problems concerning the minimum reachability probability being compared to $1$.
We conjecture that finite-memory strategies are no more powerful than memoryless strategies
for this class of problem.   
For the three other classes of qualitative reachability problems,
we have shown that the UMC and IMDP semantics coincide. 
We note that the algorithms presented in this paper require some numerical computation
(a sum and a comparison of the result with $1$ in the $\cpre$, $\apre$ and ILEC computations),
but these operations are simpler than the polynomial-time solutions for quantitative properties of
(closed) IMCs in \cite{CHK13,PLSS13}.
Similarly, the $\cpre$ and $\apre$ operators are simpler than the polynomial-time step of value iteration
used in the context of quantitative verification of \cite{HM18}.
For the IMDP semantics, our methods give directly a P-complete algorithm for 
the qualitative fragment of the temporal logic {\sc Pctl} \cite{HJ94}.
Future work could consider quantitative properties
and $\omega$-regular properties,
and applying the results to develop qualitative reachability methods for 
interval Markov decision processes 
or for higher-level formalisms such as
clock-dependent probabilistic timed automata \cite{Spr17}.

\bibliographystyle{abbrv}
\bibliography{oimcs}

\begin{thebibliography}{10}

\bibitem{BdAFK18}
C.~Baier, L.~de~Alfaro, V.~Forejt, and M.~Kwiatkowska.
\newblock Model checking probabilistic systems.
\newblock In E.~M. Clarke, T.~A. Henzinger, H.~Veith, and R.~Bloem, editors,
  {\em Handbook of Model Checking}, pages 963--999. Springer, 2018.

\bibitem{BK08}
C.~Baier and J.-P. Katoen.
\newblock {\em Principles of model checking}.
\newblock MIT Press, 2008.

\bibitem{CDLLPW11}
B.~Caillaud, B.~Delahaye, K.~G. Larsen, A.~Legay, M.~L. Pedersen, and
  A.~Wasowski.
\newblock Constraint {M}arkov chains.
\newblock {\em Theoretical Computer Science}, 412(34):4373--4404, 2011.

\bibitem{CK15}
S.~Chakraborty and J.-P. Katoen.
\newblock Model checking of open interval {M}arkov chains.
\newblock In M.~Gribaudo, D.~Manini, and A.~Remke, editors, {\em Proc. ASMTA
  2015}, volume 9081 of {\em LNCS}, pages 30--42. Springer, 2015.

\bibitem{CSH08}
K.~Chatterjee, K.~Sen, and T.~A. Henzinger.
\newblock Model-checking omega-regular properties of interval {M}arkov chains.
\newblock In R.~Amadio, editor, {\em Proc. FOSSACS 2008}, volume 4962 of {\em
  LNCS}, pages 302--317. Springer, 2008.

\bibitem{CHK13}
T.~Chen, T.~Han, and M.~Kwiatkowska.
\newblock On the complexity of model checking interval-valued discrete time
  {M}arkov chains.
\newblock {\em Information Processing Letters}, 113(7):210--216, 2013.

\bibitem{CY95}
C.~Courcoubetis and M.~Yannakakis.
\newblock The complexity of probabilistic verification.
\newblock {\em Journal of the {ACM}}, 42(4):857--907, 1995.

\bibitem{Daw04}
C.~Daws.
\newblock Symbolic and parametric model checking of discrete-time {M}arkov
  chains.
\newblock In Z.~Liu and K.~Araki, editors, {\em Proc. ICTAC 2004}, volume 3407
  of {\em LNCS}, pages 280--294. Springer, 2004.

\bibitem{deAlf97}
L.~de~Alfaro.
\newblock {\em Formal verification of probabilistic systems}.
\newblock PhD thesis, Stanford University, Department of Computer Science,
  1997.

\bibitem{deAlf99}
L.~de~Alfaro.
\newblock Computing minimum and maximum reachability times in probabilistic
  systems.
\newblock In J.~Baeten and S.~Mauw, editors, {\em Proc. CONCUR 1999}, volume
  1664 of {\em LNCS}, pages 66--81. Springer, 1999.

\bibitem{FKNP11}
V.~Forejt, M.~Kwiatkowska, G.~Norman, and D.~Parker.
\newblock Automated verification techniques for probabilistic systems.
\newblock In M.~Bernardo and V.~Issarny, editors, {\em Formal Methods for
  Eternal Networked Software Systems (SFM 2011)}, volume 6659 of {\em LNCS},
  pages 53--113. Springer, 2011.

\bibitem{HM18}
S.~Haddad and B.~Monmege.
\newblock Interval iteration algorithm for {MDPs} and {IMDPs}.
\newblock {\em Theoretical Computer Science}, 735:111--131, 2018.

\bibitem{HJ94}
H.~Hansson and B.~Jonsson.
\newblock A logic for reasoning about time and reliability.
\newblock {\em Formal Aspects of Computing}, 6(5):512--535, 1994.

\bibitem{JL91}
B.~Jonsson and K.~G. Larsen.
\newblock Specification and refinement of probabilistic processes.
\newblock In {\em Proc. LICS 1991}, pages 266--277. {IEEE} Computer Society,
  1991.

\bibitem{KU02}
I.~O. Kozine and L.~V. Utkin.
\newblock Interval-valued finite {M}arkov chains.
\newblock {\em Reliable Computing}, 8(2):97--113, 2002.

\bibitem{LMT07}
R.~Lanotte, A.~Maggiolo{-}Schettini, and A.~Troina.
\newblock Parametric probabilistic transition systems for system design and
  analysis.
\newblock {\em Formal Aspects of Computing}, 19(1):93--109, 2007.

\bibitem{PLSS13}
A.~Puggelli, W.~Li, A.~L. Sangiovanni{-}Vincentelli, and S.~A. Seshia.
\newblock Polynomial-time verification of {PCTL} properties of {MDPs} with
  convex uncertainties.
\newblock In N.~Sharygina and H.~Veith, editors, {\em Proc. CAV 2013}, volume
  8044 of {\em LNCS}, pages 527--542. Springer, 2013.

\bibitem{SVA06}
K.~Sen, M.~Viswanathan, and G.~Agha.
\newblock Model-checking {M}arkov chains in the presence of uncertainties.
\newblock In H.~Hermanns and J.~Palsberg, editors, {\em Proc. TACAS 2006},
  volume 3920 of {\em LNCS}, pages 394--410, 2006.

\bibitem{Spr17}
J.~Sproston.
\newblock Probabilistic timed automata with clock-dependent probabilities.
\newblock In M.~Hague and I.~Potapov, editors, {\em Proc. RP 2017}, volume
  10506 of {\em LNCS}, pages 144--159. Springer, 2017.

\bibitem{Var85}
M.~Vardi.
\newblock Automatic verification of probabilistic concurrent finite-state
  programs.
\newblock In {\em Proc. FOCS 1985}, pages 327--338. {IEEE} Computer Society,
  1985.

\end{thebibliography}

\section{Appendix}

\subsection{Proof of \lemref{lem:validass}}

($\Rightarrow$)
Let $\edgesubset$ be a valid subset of $\edgesfrom{\astate}$.
Firstly, because $\edgesubset$ is large,
and from well formedness,
we have that there exists some function $f: \edgesubset \ra [0,1]$
such that $f(\astate,\astate') \in \otrans(\astate,\astate')$ for each $(\astate,\astate') \in \edgesubset$
and $\sum_{(\astate,\astate') \in \edgesubset} f(\astate,\astate') = 1$.
Next, because $\edgesubset$ is realisable,
we can obtain an assignment $\apa{\astate}$ for $\astate$
by letting $\apa{\astate}(\astate') = f(\astate,\astate')$ for each $(\astate,\astate') \in \edgesubset$,
and letting $\apa{\astate}(\astate') = 0$ for each $(\astate,\astate') \not\in \edgesubset$.
This definition is clearly such that
$\set{(\astate,\astate') \mid \astate' \in \support(\apa{\astate})} = \edgesubset$.

$(\Leftarrow)$
Let $\apa{\astate}$ be an assignment for $\astate$
such that $\set{(\astate,\astate') \mid \astate' \in \support(\apa{\astate})} = \edgesubset$.
We first show that $\edgesubset$ is large.
By the definition of assignments,
we have $\apa{\astate} \in \otrans(\astate,\astate')$, 
and hence $\apa{\astate}(\astate') \leq \rep{\otrans(\astate,\astate')}$
for each state $\astate' \in \ostates$.
From this fact,
and given that $\apa{\astate}$ is a distribution (i.e., sums to $1$),
we have:
\[ 
1 = \sum_{\astate' \in \support(\apa{\astate})} \apa{\astate}(\astate')
\leq \sum_{\astate' \in \support(\apa{\astate})} \rep{\otrans(\astate,\astate')}
= \sum_{(\astate,\astate') \in \edgesubset} \rep{\otrans(\astate,\astate')} \; .
\]
In the case of $\sum_{(\astate,\astate') \in \edgesubset} \rep{\otrans(\astate,\astate')} > 1$,
we have shown that $\edgesubset$ is large.
In the case of $\sum_{(\astate,\astate') \in \edgesubset} \rep{\otrans(\astate,\astate')} = 1$,
then $\apa{\astate}(\astate') = \rep{\otrans(\astate,\astate')}$ 
(i.e., $\apa{\astate}(\astate')$ must be equal to the right endpoint of $\otrans(\astate,\astate')$)
for each $(\astate,\astate') \in \edgesubset$.
Hence $\edgesubset \subseteq \edgespara{\rc}$,
and as a consequence $\edgesubset$ is large.
The fact that $\set{(\astate,\astate') \mid \astate' \in \support(\apa{\astate})} = \edgesubset$
implies that $\apa{\astate}(\astate') = 0$ for all $(\astate,\astate') \in \edgesfrom{\astate} \setminus \edgesubset$.
Hence $\edgesfrom{\astate} \setminus \edgesubset \subseteq \edgespara{\lcz}$,
and therefore $\edgesubset$ is realisable.
Because $\edgesubset$ is large and realisable, $\edgesubset$ is valid.
\qed


\subsection{Proofs from Section \ref{sec:umc_algos}}

\subsubsection{Proof of \lemref{lem:reach_umc}.}
$(\Rightarrow)$
Let $\adtmc = (\ostates,\dtrans) \in \umc{\anoimc}$ be a DTMC 
such that $\dprob{\adtmc}{\astate}(\reachpaths{\targetset}) > 0$.
Because $\dprob{\adtmc}{\astate}(\reachpaths{\targetset}) > 0$,
there exists at least one finite path $\astate_0 \astate_1 \cdots \astate_n \in \dfinpaths{\adtmc}(\astate)$ 
such that $\astate_n \in \targetset$.
For each $i < n$, we have that $\dtrans(\astate_i,\astate_{i+1})>0$.
From this fact, and by the definition of assignment, we must have that $\otrans(\astate_i,\astate_{i+1}) \neq [0,0]$.
This means that $(\astate_i,\astate_{i+1}) \in \edges$.
Repeating this reasoning for each $i < n$,
we have that $\astate_0 \astate_1 \cdots \astate_n \in \gfinpaths{\anoimc}(\astate)$.
Recalling that $\astate_n \in \targetset$,
this direction of the proof is completed.

$(\Leftarrow)$
Let $\astate_0 \astate_1 \cdots \astate_n \in \gfinpaths{\anoimc}(\astate)$ be a finite path 
of $\anoimc$ such that $\astate_n \in \targetset$.
By the definition of finite paths of $\anoimc$,
we have $\otrans(\astate_i,\astate_{i+1}) \neq [0,0]$ for each $i < n$.
This implies that, for each $i < n$, there exists an assignment to state $\astate_i$
that sets positive probability to $\astate_{i+1}$.
In turn, this means that $\dtrans(\astate_i,\astate_{i+1})>0$ for each $i < n$, 
for some DTMC $\adtmc = (\ostates,\dtrans) \in \umc{\anoimc}$.
Given that $\astate_n \in \targetset$,
hence we have $\dprob{\adtmc}{\astate}(\reachpaths{\targetset}) > 0$.
\qed

\subsubsection{Proof of \lemref{lem:qa-umc}.}
Before we present the proof of \lemref{lem:qa-umc}, 
we recall (and specialise to reachability) a technical result from \cite{CSH08}
that specifies that the same qualitative properties are satisfied
on (finite) DTMCs that have the same graph.
Two finite DTMCs $\adtmc_1 = (\dstates_1,\dtrans_1)$ and $\adtmc_2 = (\dstates_1,\dtrans_2)$
are \emph{graph equivalent}
if (1) $\dstates_1 = \dstates_2$ and
(2) $\set{(\astate,\astate') \mid \dtrans_1(\astate,\astate')>0} 
= \set{(\astate,\astate') \mid \dtrans_2(\astate,\astate')>0}$.

\begin{lemma}[\cite{CSH08}]\label{lem:graphequiv}
Given graph equivalent finite DTMCs 
$\adtmc_1 = (\dstates_1,\dtrans_1)$ and $\adtmc_2 = (\dstates_2,\dtrans_2)$,
for any state $\astate \in \dstates_1 (=\dstates_2)$, $\bowtie \in \set{<, =, >}$ and $\lambda \in \set{0,1}$,
we have $\dprob{\adtmc_1}{\astate}(\reachpaths{\targetset}) \bowtie \lambda$
if and only if $\dprob{\adtmc_2}{\astate}(\reachpaths{\targetset}) \bowtie \lambda$.
\end{lemma}

$(\Rightarrow)$
Let $\adtmc = (\ostates,\dtrans)$ be a DTMC such that $\adtmc \in \umc{\anoimc}$ 
and $\dprob{\adtmc}{\astate}(\reachpaths{\targetset}) = 0$.
Then we define the (memoryless) scheduler $\aschedpara{\adtmc} \in \scheds{\qa{\anoimc}{\waf}}$ of $\qa{\anoimc}{\waf}$
in the following way.
Consider a state $\astate' \in \ostates$,
and let $\edgesubsetpara{\astate'} = \set{ (\astate',\astate'') \in \edges \mid \dtrans(\astate',\astate'')>0 }$
be the set of edges assigned positive probability by $\adtmc$.
Note that $\edgesubsetpara{\astate'} \in \validsetsfrom{\astate'}$:
this follows from the fact that $\dtrans(\astate', \cdot)$ is an assignment for $\astate'$
and by \lemref{lem:validass}.
Then, for finite path $\afinpath \in \mfinpaths{\qa{\anoimc}{\waf}}(\astate)$
ending in state $\astate'$ (that is, $\last(\afinpath) = \astate'$),
we let $\aschedpara{\adtmc}(\afinpath) = \dirac{\waf(\edgesubsetpara{\astate'})}$,
i.e., $\aschedpara{\adtmc}$ chooses (with probability 1) the distribution 
that corresponds to the witness assignment function
applied to the edge set $\edgesubsetpara{\astate'}$
(this is possible because 
$\qatrans{\waf}(\astate') = \set{ \waf(\edgesubset) \mid \edgesubset \in \validsetsfrom{\astate'} }$ by definition).
The fact that $\dprob{\adtmc}{\astate}(\reachpaths{\targetset}) \bowtie \lambda$
implies $\mprob{\aschedpara{\adtmc}}{\astate}(\reachpaths{\targetset}) \bowtie \lambda$
then follows from the fact that 
the folded DTMC $\afolddtmcpara{\aschedpara{\adtmc}}$ of $\aschedpara{\adtmc}$
and the DTMC $\adtmc$ are graph equivalent
(because, for all states $\astate' \in \ostates$, we have
$\set{ (\astate',\astate'') \in \edges \mid \dtrans(\astate',\astate'')>0 } 
= \edgesubsetpara{\astate'} 
= \set{ (\astate',\astate'') \mid \astate' \in \support(\waf(\edgesubsetpara{\astate'})) }$),
and from \lemref{lem:graphequiv}.

$(\Leftarrow)$
Assume that there exists a scheduler $\asched \in \scheds{\qa{\anoimc}{\waf}}$ of $\qa{\anoimc}{\waf}$
such that $\mprob{\asched}{\astate}(\reachpaths{\targetset}) = 0$.
Given that $\qa{\anoimc}{\waf}$ is a finite MDP,
from standard results \cite{deAlf97},
we can assume that $\asched$ is (a) pure
(that is, for all finite paths $\afinpath \in \mfinpaths{\qa{\anoimc}{\waf}}$,
we have that $\asched(\afinpath) = \dirac{\adist}$ for some $\adist \in \qatrans{\waf}(\last(\afinpath))$)
and (b) memoryless.
Then we consider the DTMC $\adtmc = (\ostates,\dtrans)$, 
where, 
for all states $\astate' \in \ostates$, we define $\dtrans(\astate',\cdot) = \asched(\astate')$.
Given that $\asched(\astate')$ is an assignment
(because $\asched$ is pure, i.e., 
assigns probability 1 to a distribution $\waf(\edgesubset)$, for some $\edgesubset \in \validsetsfrom{\astate'}$,
and because $\waf$ is a witness assignment funcion,
i.e., $\waf(\edgesubset)$ is an assignment), 
we have that $\adtmc \in \umc{\anoimc}$.
The folded DTMC $\afolddtmcpara{\asched}$ of $\asched$ and $\adtmc$ are identical.
Hence, given that $\mprob{\asched}{\astate}(\reachpaths{\targetset}) \bowtie \lambda$,
we have $\dprob{\adtmc}{\astate}(\reachpaths{\targetset}) \bowtie \lambda$.
\qed

\subsubsection{Proof of \lemref{lem:cpre_eff}.}
We start by recalling the definition of $\cpre$:
\[
\begin{array}{clr}
\multicolumn{3}{l}
{\astate \in \cpre(\astateset)}
\\
\Leftrightarrow &
\exists \adist \in \qatrans{\waf}(\astate) \ldot \support(\adist) \subseteq \astateset
\\
\Leftrightarrow &
\exists \edgesubset \in \edgesfrom{\astate} \ldot \edgesubset \subseteq \ostates \times \astateset 
\wedge \edgesubset \in \validsetsfrom{\astate}
& \mbox{(definition of $\qa{\anoimc}{\waf}$)}
\\
\Leftrightarrow &
\exists \edgesubset \in \edgesfrom{\astate} \ldot \edgesubset \subseteq \ostates \times \astateset 
\wedge \edgesfrom{\astate} \setminus \edgesubset \subseteq \edgespara{\lcz}
\wedge \edgesubset \mbox{ is large}
&
\\
\multicolumn{3}{r}
{\mbox{(definition of $\validsetsfrom{\astate}$)}}
\\
\Leftrightarrow &
\multicolumn{2}{l}
{
\edgesfromtwo{\astate}{\astateset} \subseteq \ostates \times \astateset 
\wedge \edgesfrom{\astate} \setminus \edgesfromtwo{\astate}{\astateset} 
\subseteq \edgespara{\lcz}
\wedge \edgesfromtwo{\astate}{\astateset} \mbox{ is large}
}
\\
\multicolumn{3}{r}
{\mbox{(subsituting $\edgesfromtwo{\astate}{\astateset}$ for $\edgesubset$)}}
\\
\Leftrightarrow &
\edgesfromtwo{\astate}{\ostates \setminus \astateset} \subseteq \edgespara{\lcz}
\wedge \edgesfromtwo{\astate}{\astateset} \mbox{ is large}
&
\end{array}
\]
where the last step follows from the fact that 
$\edgesfromtwo{\astate}{\astateset} \subseteq \ostates \times \astateset$
holds trivially,
and from $\edgesfrom{\astate} \setminus \edgesfromtwo{\astate}{\astateset}
= \edgesfromtwo{\astate}{\ostates \setminus \astateset}$.
\qed

\subsubsection{Proof of \lemref{lem:apre_eff}.}
From the definition of $\apre$:
\[
\begin{array}{cl}
\multicolumn{2}{l}
{\astate \in \apre(\anotherstateset, \astateset)}
\\
\Leftrightarrow &
\exists \adist \in \qatrans{\waf}(\astate) \ldot 
\support(\adist) \subseteq \anotherstateset \wedge \support(\adist) \cap \astateset \neq \emptyset
\\
\Leftrightarrow &
\exists \edgesubset \in \edgesfrom{\astate} \ldot 
\edgesubset \subseteq \ostates \times \anotherstateset
\wedge \edgesubset \cap (\ostates \times \astateset) \neq \emptyset
\wedge \edgesubset \in \validsetsfrom{\astate}
\hspace{2.7cm}
\\
\multicolumn{2}{r}
{\mbox{(definition of $\qa{\anoimc}{\waf}$)}}
\\
\end{array}
\]

\[
\begin{array}{cl}
\Leftrightarrow &
\exists \edgesubset \in \edgesfrom{\astate} \ldot 
\edgesubset \subseteq \ostates \times \anotherstateset
\wedge \edgesubset \cap (\ostates \times \astateset) \neq \emptyset
\wedge \edgesfrom{\astate} \setminus \edgesubset \subseteq \edgespara{\lcz}
\wedge \edgesubset \mbox{ is large}
\\
\multicolumn{2}{r}
{\mbox{(definition of $\validsetsfrom{\astate}$)}}
\\
\Leftrightarrow &
\edgesfromtwo{\astate}{\anotherstateset} \subseteq \ostates \times \anotherstateset
\wedge \edgesfromtwo{\astate}{\anotherstateset} \cap (\ostates \times \astateset) \neq \emptyset
\wedge \edgesfrom{\astate} \setminus \edgesfromtwo{\astate}{\anotherstateset} \subseteq \edgespara{\lcz}
\\
&
\wedge \edgesfromtwo{\astate}{\anotherstateset} \mbox{ is large}
\\
\multicolumn{2}{r}
{\mbox{(subsituting $\edgesfromtwo{\astate}{\anotherstateset}$ for  $\edgesubset$)}}
\\
\Leftrightarrow &
\edgesfromtwo{\astate}{\astateset \cap \anotherstateset} \neq \emptyset
\wedge \edgesfromtwo{\astate}{\ostates \setminus \anotherstateset} \subseteq \edgespara{\lcz}
\wedge \edgesfromtwo{\astate}{\anotherstateset} \mbox{ is large}
\end{array}
\]
where the last step follows from the fact that 
$\edgesfromtwo{\astate}{\anotherstateset} \subseteq \ostates \times \anotherstateset$
holds trivially,
from $\edgesfromtwo{\astate}{\anotherstateset} \cap (\ostates \times \astateset)
= \edgesfromtwo{\astate}{\astateset \cap \anotherstateset}$,
and from $\edgesfrom{\astate} \setminus (\edgesfromtwo{\astate}{\anotherstateset})
= \edgesfromtwo{\astate}{\ostates \setminus \anotherstateset}$.
\qed

\subsubsection{Proof of \lemref{lem:min_one_umc}.}
$(\Rightarrow)$
Let $\adtmc = (\ostates,\dtrans) \in \umc{\anoimc}$ be a DTMC 
such that $\dprob{\adtmc}{\astate}(\reachpaths{\targetset}) < 1$.
A bottom strongly connected component (BSCC) $\abscc \subseteq \ostates$ of $\adtmc$
is a strongly connected component of the graph of $\adtmc$
(that is, the graph $(\ostates,\set{(\astate',\astate'') \mid \dtrans(\astate',\astate'')>0})$)
such that there is no outgoing edge from $\abscc$
(that is, for all states $\astate' \in \abscc$ and $\astate'' \in \ostates$, 
if $\dtrans(\astate',\astate'')>0$ then $\astate'' \in \abscc$).
Let $\bsccset \subseteq 2^\ostates$ be the set of BSCCs of $\adtmc$.
Given that $\adtmc$ is a finite DTMC,
by standard results for finite DTMCs, we have that BSCCs are reached with probability $1$,
i.e., $\dprob{\adtmc}{\astate}(\reachpaths{\bigcup_{\abscc \in \bsccset} \abscc}) = 1$,
and that once a BSCC is entered all of its states are visited with probability $1$.
Hence $\dprob{\adtmc}{\astate}(\reachpaths{\targetset}) < 1$
implies that there exists some BSCC $\abscc \in \bsccset$ such that $\abscc \cap \targetset = \emptyset$
and $\dprob{\adtmc}{\astate}(\reachpaths{\abscc}) > 0$.
At this point we repeat the reasoning of \lemref{lem:reach_umc}
to show that there exists a finite path $\astate_0 \astate_1 \cdots \astate_n \in \gfinpaths{\anoimc}(\astate)$
such that $\astate_n \in \abscc$.
Now we show that $\abscc \subseteq \eqset{0}{\uident}$:
for any state $\astate' \in \abscc$,
the fact that $\abscc \cap \targetset = \emptyset$ and from the fact that BSCCs do not feature outgoing edges,
we have that $\dprob{\adtmc}{\astate'}(\reachpaths{\targetset}) = 0$,
and hence $\astate' \in \eqset{0}{\uident}$.
Therefore, for the finite path $\astate_0 \astate_1 \cdots \astate_n \in \gfinpaths{\anoimc}(\astate)$,
we have $\astate_n \in \eqset{0}{\uident}$,
which completes this direction of the proof.

$(\Leftarrow)$
Let $\astate_0 \astate_1 \cdots \astate_n \in \gfinpaths{\anoimc}(\astate)$ be a finite path 
of $\anoimc$ such that $\astate_n \in \eqset{0}{\uident}$.
We assume w.l.o.g. that $\astate_i \not\in \eqset{0}{\uident}$ for all $i < n$.
As in the proof of \lemref{lem:reach_umc},
we can use this fact to conclude that there exists a DTMC $\adtmc = (\ostates,\dtrans) \in \umc{\anoimc}$
such that $\dtrans(\astate_i,\astate_{i+1})>0$ for each $i < n$.
Furthermore, for any state $\astate' \in \eqset{0}{\uident}$,
there exists a DTMC $\adtmcparaone{\astate'} = (\ostates,\dtranspara{\astate'}) \in \umc{\anoimc}$
such that $\dprob{\adtmcparaone{\astate'}}{\astate'}(\reachpaths{\targetset}) = 0$.
From $\astate_n \in \eqset{0}{\uident}$, 
we have $\dprob{\adtmcparaone{\astate_n}}{\astate_n}(\reachpaths{\targetset}) = 0$.
Note that all states on all paths in $\dfinpaths{\adtmcparaone{\astate_n}}(\astate_n)$
are in $\eqset{0}{\uident}$
(if this was not the case, that is there exists a path in $\dfinpaths{\adtmcparaone{\astate_n}}(\astate_n)$
featuring a state $\astate'' \not\in \eqset{0}{\uident}$,
then we must also have $\astate_n \not\in \eqset{0}{\uident}$,
i.e., we have obtained a contradiction).
The fact that the set $\set{ \astate_0, \astate_1, \cdots, \astate_{n-1} }$ 
(i.e., the states before $\astate_n$ along the finite path)
and the set $\eqset{0}{\uident}$ are disjoint means that we
can combine the DTMCs $\adtmc$ along the finite path 
and $\adtmcparaone{\astate_n}$ within $\eqset{0}{\uident}$ to obtain a DTMC $\adtmc' = (\ostates,\dtrans')$;
formally, $\dtrans'(\astate_i,\cdot) =  \dtrans(\astate_i,\cdot)$ for $i < n$,
$\dtrans'(\astate',\cdot) =  \dtranspara{\astate_n}(\astate',\cdot)$ for $\astate' \in \eqset{0}{\uident}$,
and $\dtrans'(\astate',\cdot)$ can be defined by arbitrary assignments for all other states, 
i.e., $\astate' \in \ostates \setminus (\set{ \astate_0, \astate_1, \cdots, \astate_{n-1} } \cup \eqset{0}{\uident})$.
Given that $\adtmc, \adtmcparaone{\astate_n} \in \umc{\anoimc}$,
we have that $\adtmc' \in \umc{\anoimc}$.
Furthermore, in $\adtmc'$, from $\astate$ there exists a finite path to states $\eqset{0}{\uident}$
in which the DTMC is subsequently confined, 
and (by the trivial fact that $\eqset{0}{\uident} \subseteq \ostates \setminus \targetset$) 
hence $\dprob{\adtmc'}{\astate}(\reachpaths{\targetset}) < 1$,
completing this direction of the proof.
\qed


\subsection{Proofs from Section \ref{sec:imdp_algos}}

\subsubsection{Proof of \lemref{lem:reach_imc}.}
$(\Rightarrow)$
Let $\asched \in \scheds{\imdp{\anoimc}}$ be a scheduler of $\imdp{\anoimc}$ 
such that $\mprob{\asched}{\astate}(\reachpaths{\targetset}) > 0$.
Recall that the DTMC $\adtmcparatwo{\asched}{\astate}$ is defined as $(\mfinpaths{\asched}(\astate),\dtrans)$,
i.e., the state set comprises the finite paths resulting from choices of $\asched$ from state $\astate$,
and, for finite paths $\afinpath,\afinpath' \in \mfinpaths{\asched}(\astate)$,
we have $\dtrans(\afinpath,\afinpath')>0$ if $\afinpath' = \afinpath \adist \astate'$
such that $\adist \in \mtrans(\last(\afinpath))$, $\asched(\adist)>0$ and $\adist(\astate')>0$,
otherwise $\dtrans(\afinpath,\afinpath')=0$.
Furthermore, we say that a path $\apath$ of $\adtmcparatwo{\asched}{\astate}$ is in $\reachpaths{\targetset}$
if the last state of a finite path that is a prefix of $\apath$ is in $\targetset$.
For a path $\apath$ of $\adtmcparatwo{\asched}{\astate}$,
we can obtain a path $\apath'$ of $\anoimc$ simply by extracting 
the sequence of the final states for all finite prefixes of $\apath$:
that is, from the path 
$\astate_0 (\astate_0 \adist_0 \astate_1) (\astate_0 \adist_0 \astate_1 \adist_1 \astate_2) \cdots$ 
of $\adtmcparatwo{\asched}{\astate}$
we can obtain the path $\astate_0 \astate_1 \astate_2 \cdots$ of $\anoimc$,
because $\adist_i(\astate_{i+1})>0$
implies that $\otrans(\astate_i,\astate_{i+1}) \neq [0,0]$
and hence $(\astate_i,\astate_{i+1}) \in \edges$ for all $i \in \Nset$.
Observe that $\apath \in \reachpaths{\targetset}$ implies that $\apath' \in \reachpaths{\targetset}$.
Hence $\mprob{\asched}{\astate}(\reachpaths{\targetset}) > 0$
implies that there exists $\apath \in \dpaths{\adtmcparatwo{\asched}{\astate}}(\astate)$ 
such that $\apath \in \reachpaths{\targetset}$,
which in turn implies that there exists $\apath' \in \gpaths{\anoimc}(\astate)$ 
such that $\apath' \in \reachpaths{\targetset}$,
i.e., there exists a finite prefix $\afinpath$ of $\apath$ 
(and therefore $\afinpath \in \gfinpaths{\anoimc}(\astate)$)
such that $\last(\afinpath) \in \targetset$.

$(\Leftarrow)$
Let $\astate_0 \astate_1 \cdots \astate_n \in \gfinpaths{\anoimc}(\astate)$ be a finite path 
of $\anoimc$ such that $\astate_n \in \targetset$.
By the definition of finite paths of $\anoimc$,
we have $\otrans(\astate_i,\astate_{i+1}) \neq [0,0]$ for each $i < n$,
which implies in turn that there exists an assignment $\apa{}_i$ to $\astate_i$ 
such that $\apa{}_i(\astate_{i+1})>0$.
We consider a scheduler $\asched \in \scheds{\imdp{\anoimc}}$ of $\imdp{\anoimc}$
such that, for any finite path $\afinpath \in \mfinpaths{\imdp{\anoimc}}(\astate)$,
if $\last(\afinpath) = \astate_i$ then $\asched(\afinpath) = \apa{}_i$ for $i < n$
($\asched$ can be defined in an arbitrary manner for all other finite paths).
Observe that the finite path 
$\afinpath = \astate_0 (\astate_0 \apa{}_0 \astate_1) (\astate_0 \apa{}_0 \astate_1 \apa{}_1 \astate_2)
\cdots (\astate_0 \apa{}_0 \cdots \apa{}_{n-1} \astate_n)$
is a finite path of $\adtmcparatwo{\asched}{\astate}$ 
such that all infinite paths that have $\afinpath$ as a prefix are in $\reachpaths{\targetset}$,
and hence $\mprob{\asched}{\astate}(\reachpaths{\targetset}) > 0$.
\qed

\subsubsection{Proof of \lemref{lem:exists_imdp}.}
The proof consists of two parts, depending on whether $\lambda = 0$ or $\lambda = 1$.

{\sl Case: $\lambda = 0$.}
$(\Rightarrow)$
Let $\asched \in \scheds{\imdp{\anoimc}}$ be a scheduler of $\imdp{\anoimc}$
such that $\mprob{\asched}{\astate}(\reachpaths{\targetset}) = 0$.
We show how we can define a scheduler $\vsa{\asched} \in \scheds{\qa{\anoimc}{\waf}}$
such that $\mprob{\vsa{\asched}}{\astate}(\reachpaths{\targetset}) = 0$. 

For any finite path $\afinpath \in \mfinpaths{\imdp{\anoimc}}(\astate)$,
let $\edgesubsetparatwo{\asched}{\afinpath}$ be the set of edges assigned positive probability 
by $\asched$ after $\afinpath$.
Formally: 
\[
\edgesubsetparatwo{\asched}{\afinpath} = 
\set{ (\astate,\astate') \in \edges \mid \astate = \last(\afinpath) 
\wedge \exists \adist \in \mtrans(\astate) \ldot (\asched(\afinpath)(\adist)>0 \wedge \adist(\astate')>0) } \, .
\]
We now define the partial function 
$\asvnopara : \mfinpaths{\qa{\anoimc}{\waf}}(\astate) \ra \mfinpaths{\asched}(\astate)$,
which associates with a finite path of $\qa{\anoimc}{\waf}$ some finite path of $\imdp{\anoimc}$ 
(more precisely, of $\asched$)
that features the same sequence of states.
For $\afinpath = \astate_0 \adist_0 \astate_1 \adist_1 \cdots \adist_{n-1} \astate_n
\in \mfinpaths{\qa{\anoimc}{\waf}}(\astate)$,
let $\asv{\afinpath} = \astate_0' \adist_0' \astate_1' \adist_1' \cdots  \adist_{n-1}' \astate_n'$
such that:
\begin{itemize}
\item
$\astate_i' = \astate_i$ for all $i \leq n$; 
\item
$\asched(\asv{\astate_0 \adist_0 \astate_1 \adist_1 \cdots \adist_{i-1} \astate_i})(\adist_i')>0$ 
and $\adist_i'(\astate_{i+1}')>0$ 
for all $i < n$;
\item
$\asv{\astate_0 \adist_0 \astate_1 \adist_1 \cdots \adist_{i-1} \astate_i} 
= \astate_0' \adist_0' \astate_1' \adist_1' \cdots  \adist_{i-1}' \astate_i'$ for all $i < n$.
\end{itemize}
The first two conditions and $\astate_0 = \astate$ 
ensure that $\asv{\afinpath} \in \mfinpaths{\asched}(\astate)$;
if these conditions do not hold, then $\asv{\afinpath}$ is undefined.
The third condition ensures that $\asvnopara$ maps paths of $\mfinpaths{\qa{\anoimc}{\waf}}(\astate)$ 
to paths of $\mfinpaths{\asched}(\astate)$ in a consistent manner 
(for example, the condition ensures that, if there are two finite paths $\afinpath_1$ and $\afinpath_2$
of $\qa{\anoimc}{\waf}$ with a common prefix of length $i$,
then $\asv{\afinpath_1}$ and $\asv{\afinpath_2}$ will also have a common prefix of length $i$).
We are now in a position to define the scheduler $\vsa{\asched}$ of $\qa{\anoimc}{\waf}$:
let $\vsa{\asched}(\afinpath) = \dirac{\waf(\edgesubsetparatwo{\asched}{\asv{\afinpath}})}$ 
for each finite path $\afinpath \in \mpaths{\qa{\anoimc}{\waf}}(\astate)$.

Next we need to show that $\mprob{\vsa{\asched}}{\astate}(\reachpaths{\targetset}) = 0$.
Note that $\mprob{\asched}{\astate}(\reachpaths{\targetset}) = 0$
if and only if all paths $\apath \in \mpaths{\asched}(\astate)$ are 
such that $\apath \not\in \reachpaths{\targetset}$,
and similarly $\mprob{\vsa{\asched}}{\astate}(\reachpaths{\targetset}) = 0$
if and only if all paths $\apath \in \mpaths{\vsa{\asched}}(\astate)$ are
such that $\apath \not\in \reachpaths{\targetset}$.
Hence we show that the existence of a path $\apath \in \mpaths{\vsa{\asched}}(\astate)$
such that $\apath \in \reachpaths{\targetset}$
implies the existence of a path $\apath' \in \mpaths{\asched}(\astate)$
such that $\apath' \in \reachpaths{\targetset}$.
Let $\apath \in \mpaths{\vsa{\asched}}(\astate)$ be
such that $\apath \in \reachpaths{\targetset}$.
We can identify a path $\apath' \in \mpaths{\asched}(\astate)$
that visits the same states as $\apath$, 
and hence is such that $\apath' \in \reachpaths{\targetset}$.
Let $\apath = \astate_0 \adist_0 \astate_1 \adist_1 \cdots$,
and let $\afinpathpref{i}$ be the $i$-th prefix of $\apath$
(i.e., $\afinpathpref{i} = \astate_0 \adist_0 \astate_1 \adist_1 \cdots \adist_{i-1} \astate_i$).
We show the existence of $\apath'$ by induction, 
by considering prefixes $\afinpathpref{i}'$ of $\apath'$ of increasing length.
\begin{description}
\item[{\sl (Base case.)}]
Let $\afinpathpref{0}' = \astate_0$.
\item[{\sl (Inductive step.)}]
\sloppypar{
Assume that we have constructed the finite path 
$\afinpathpref{i}' = \astate_0 \adist_0' \astate_1 \adist_1' \cdots \adist_{i-1}' \astate_i$.
We show how to extend $\afinpathpref{i}'$ with one transition to obtain the finite path $\afinpathpref{i+1}'$.
The transition $\astate_i \adist_i \astate_{i+1}$ along $\apath$ implies
that $(\astate_i,\astate_{i+1}) \in \edgesubsetparatwo{\asched}{\asv{\afinpathpref{i}}}$.
This fact then implies that there exists some $\adist' \in \mtrans(\astate_i)$
such that $\asched(\asv{\afinpathpref{i}})(\adist')>0$ and $\adist'(\astate_{i+1})>0$.
We then let $\afinpathpref{i+1}' = \afinpathpref{i}' \adist' \astate_{i+1}$.
}
\end{description}
From this construction of $\apath'$,
we can see that $\apath' \in \mpaths{\asched}(\astate)$
(because $\apath'$ follows the definition of a path of scheduler $\asched$,
and because the choice of $\adist'$ after $\afinpathpref{i}'$
must be consistent with the choices made along all prefixes of $\afinpathpref{i}'$).
Furthermore, we have $\apath' \in \reachpaths{\targetset}$
(because $\apath$ and $\apath'$ feature the same states, and $\apath \in \reachpaths{\targetset}$).
Hence we have shown that 
$\mprob{\asched}{\astate}(\reachpaths{\targetset}) = 0$ implies
$\mprob{\vsa{\asched}}{\astate}(\reachpaths{\targetset}) = 0$,
concluding this direction of the proof.

$(\Leftarrow)$
Let $\asched \in \scheds{\qa{\anoimc}{\waf}}$ 
such that $\mprob{\asched}{\astate}(\reachpaths{\targetset}) = 0$.
Given that $\qa{\anoimc}{\waf}$ is a finite MDP, 
we can assume that $\asched$ is memoryless and pure.
We now define $\asched' \in \scheds{\imdp{\anoimc}}$ 
and show that $\mprob{\asched'}{\astate}(\reachpaths{\targetset}) = 0$.
For a finite path $\afinpath \in \mfinpaths{\imdp{\anoimc}}(\astate)$,
let $\asched'(\afinpath) = \asched(\last(\afinpath))$
(recall that, for all states $\astate' \in \ostates$,
$\asched(\astate') = \dirac{\waf(\edgesubset)}$ for some $\edgesubset \in \validsetsfrom{\astate'}$,
where $\waf(\edgesubset)$ is an assignment for $\astate'$,
hence $\asched'$ is well defined).
Then we have that the DTMCs $\adtmcparatwo{\asched}{\astate}$ and $\adtmcparatwo{\asched'}{\astate}$
are identical,
and hence $\mprob{\asched}{\astate}(\reachpaths{\targetset}) = 0$ implies that 
$\mprob{\asched'}{\astate}(\reachpaths{\targetset}) = 0$.

Hence the part of the proof for $\lambda = 0$ is concluded.

{\sl Case: $\lambda = 1$.}
$(\Rightarrow)$
Let $\asched \in \scheds{\imdp{\anoimc}}$ be such that $\mprob{\asched}{\astate}(\reachpaths{\targetset}) = 1$.
We use the construction of the scheduler $\vsa{\asched}$ from the case of $\eqset{0}{\iident}$
(i..e, the case $\lambda = 0$ of this proof),
and show that $\mprob{\vsa{\asched}}{\astate}(\reachpaths{\targetset}) = 1$.
We proceed to show that $\mprob{\asched}{\astate}(\reachpaths{\targetset}) = 1$
implies $\mprob{\vsa{\asched}}{\astate}(\reachpaths{\targetset}) = 1$
by showing the contrapositive,
i.e., $\mprob{\vsa{\asched}}{\astate}(\reachpaths{\targetset}) < 1$
implies $\mprob{\asched}{\astate}(\reachpaths{\targetset}) < 1$.
To show this property, we use properties of \emph{end components}.
An end component of finite MDP $\qa{\anoimc}{\waf} = (\ostates,\qatrans{\waf})$
is a pair $(\ecstates,\ecdists)$ where $\ecstates \subseteq \ostates$ 
and $\ecdists: \ecstates \ra 2^{\dist(\ostates)}$ is such that
(1) $\emptyset \neq \ecdists(\astate') \subseteq \qatrans{\waf}(\astate')$
for all $\astate' \in \ecstates$,
(2) $\support(\adist) \subseteq \ecstates$ for all $\astate' \in \ecstates$ and $\adist \in \ecdists(\astate')$, and
(3) the graph 
$(\ecstates,\set{ (\astate',\astate'') \in \ecstates \times \ecstates \mid 
\exists \adist \in \ecdists(\astate') \ldot \adist(\astate'')>0 })$ 
is strongly connected.
Let $\ecs$ be the set of end components of $\qa{\anoimc}{\waf}$.
For an end component $(\ecstates,\ecdists)$,
let $\sa{(\ecstates,\ecdists)} = \set{ (\astate',\adist) \mid \astate \in \ecstates \wedge \adist \in \ecdists(\astate') }$
be the set of state-action pairs associated with $(\ecstates,\ecdists)$.
For a path $\apath \in \mpaths{\qa{\anoimc}{\waf}}(\astate)$,
we let $\infalong{\apath} = \set{ (\astate,\adist) \mid \astate \mbox{ and } \adist \mbox{ appear infinitely often along } \apath}$.
The fundamental theorem of end components \cite{deAlf97} specifies that, 
for any scheduler $\asched' \in \scheds{\qa{\anoimc}{\waf}}$,
we have that 
$\mprob{\asched'}{\astate}(\set{ \apath \mid 
\exists (\ecstates,\ecdists) \in \ecs \ldot \infalong{\apath} = \sa{(\ecstates,\ecdists)}}) = 1$.

Now, given that $\qa{\anoimc}{\waf}$ is a finite MDP,
and from $\mprob{\vsa{\asched}}{\astate}(\reachpaths{\targetset}) < 1$
and the fact that states in $\targetset$ are absorbing,
there exists an end component $(\ecstates,\ecdists) \in \ecs$
such that $\ecstates \cap \targetset = \emptyset$ and 
$\mprob{\vsa{\asched}}{\astate}(\set{ \apath \mid \infalong{\apath} = \sa{(\ecstates,\ecdists)}}) > 0$.
We observe the following property:
for any finite path $\afinpath \in \mfinpaths{\vsa{\asched}}(\astate)$,
we have that $\set{ (\astate',\astate'') \mid \waf(\edgesubsetparatwo{\asched}{\asv{\afinpath}})(\astate'')>0 } 
= \edgesubsetparatwo{\asched}{\asv{\afinpath}}$.
From the definition of $\edgesubsetparatwo{\asched}{\asv{\afinpath}}$,
we can see that $\asched$ assigns positive probability only to those distributions that assign positive probability 
to states that are targets of edges in $\edgesubsetparatwo{\asched}{\asv{\afinpath}}$.
Putting these two facts together, 
we conclude that, for any set $\astateset \subseteq \ostates$,
we have $\support(\waf(\edgesubsetparatwo{\asched}{\asv{\afinpath}})) \subseteq \astateset$
if and only if  
$\set{ \astate' \in \ostates \mid 
\exists \adist \in \mtrans(\last(\asv{\apath})) \ldot (\asched(\asv{\afinpath})(\adist)>0 \wedge \adist(\astate')>0) } 
\subseteq \astateset$.
This means that the existence of a finite path $\afinpath \in \mfinpaths{\vsa{\asched}}(\astate)$
such that all suffixes of $\afinpath$ generated by $\vsa{\asched}$ visit only states in $\ecstates$
implies that all suffixes of $\asv{\afinpath}$ 
(which we recall is a finite path of $\asched$, i.e., $\asv{\afinpath} \in \mfinpaths{\asched}(\astate)$)
generated by $\asched$ visit only states in $\ecstates$.
Given that 
$\mprob{\vsa{\asched}}{\astate}(\set{ \apath \mid \infalong{\apath} = \sa{(\ecstates,\ecdists)}}) > 0$,
such a finite path in $\mfinpaths{\vsa{\asched}}(\astate)$ exists.
Then, given that there exists a finite path in $\mfinpaths{\asched}(\astate)$
such that all suffixes generated by $\asched$ visit only states in $\ecstates$,
and from the fact that $\ecstates \cap \targetset = \emptyset$,
we have that $\mprob{\asched}{\astate}(\reachpaths{\targetset}) < 1$.
Hence we have shown that $\mprob{\vsa{\asched}}{\astate}(\reachpaths{\targetset}) < 1$
implies that $\mprob{\asched}{\astate}(\reachpaths{\targetset}) < 1$.
This means that $\mprob{\asched}{\astate}(\reachpaths{\targetset}) = 1$
implies that $\mprob{\vsa{\asched}}{\astate}(\reachpaths{\targetset}) = 1$.
Hence, for $\asched \in \scheds{\imdp{\anoimc}}$ such that $\mprob{\asched}{\astate}(\reachpaths{\targetset}) = 1$
there exists $\asched' \in \scheds{\qa{\anoimc}{\waf}}$ 
such that $\mprob{\asched'}{\astate}(\reachpaths{\targetset}) = 1$.

$(\Leftarrow)$
Let $\asched \in \scheds{\qa{\anoimc}{\waf}}$ be
such that $\mprob{\asched}{\astate}(\reachpaths{\targetset}) = 1$. 
As in the case of the analogous direction of the case for $\lambda = 0$,
we can show the existence of $\asched' \in \scheds{\imdp{\anoimc}}$ such that 
$\adtmcparatwo{\asched}{\astate}$ and $\adtmcparatwo{\asched'}{\astate}$
are identical,
and hence $\mprob{\asched}{\astate}(\reachpaths{\targetset}) = 1$ implies that 
$\mprob{\asched'}{\astate}(\reachpaths{\targetset}) = 1$.
\qed

\subsubsection{Proof of \lemref{lem:fund_ilecs}.}
The proof is structured in the same manner as that for classical end components in \cite{deAlf97}.
Consider $\ilecstates \subseteq \ostates$ such that $\ilecstates \not\in \ilecs$.
Our aim is to show that 
$\mprob{\asched}{\astate}(\set{ \apath \mid \infstatesalong{\apath} = \ilecstates}) = 0$.
Given that $\ilecs$ is a finite set, the required result then follows from this.

First suppose that the condition (1) in the definition of ILECs does not hold,
i.e., there exists $(\astate',\astate'') \in \edgesfrom{\astate}$ such that 
$\astate' \in \ilecstates$, $(\astate',\astate'') \in \edgespara{\lnz}$ and $\astate'' \not\in \ilecstates$.
Observe that $\lep{\otrans(\astate',\astate'')}>0$,
which means that the probability of remaining in $\ilecstates$ when visiting $\astate'$,
i.e., $1-\lep{\otrans(\astate',\astate'')}$, is strictly less than $1$. 
Given that $\astate' \in \infstatesalong{\apath}$ 
for every $\apath$ such that $\infstatesalong{\apath} = \ilecstates$,
we have that $\mprob{\asched}{\astate}(\set{ \apath \mid \infstatesalong{\apath} = \ilecstates}) 
\leq \lim_{k \ra \infty} (1-\lep{\otrans(\astate',\astate'')})^k
= 0$.

Suppose that condition (2) in the definition of ILECs does not hold,
i.e., there exists $\astate' \in \ilecstates$ such that 
$\sum_{\anedge \in \edgesfromtwo{\astate'}{\ilecstates}} \rep{\otrans(\anedge)} < 1$.
Therefore, for all $\adist \in \mtrans(\astate')$,
we must have 
$\sum_{\astate'' \in \ilecstates} \adist(\astate'') 
\leq
\sum_{\anedge \in \edgesfromtwo{\astate'}{\ilecstates}} \rep{\otrans(\anedge)} < 1$.
Hence, the probability of remaining in $\ilecstates$ when visiting $\astate'$ is strictly less than $1$.
Then, as in the case of the first condition in the definition of ILECs,
we conclude that $\mprob{\asched}{\astate}(\set{ \apath \mid \infstatesalong{\apath} = \ilecstates}) 
\leq \lim_{k \ra \infty} (\sum_{\anedge \in \edgesfromtwo{\astate'}{\ilecstates}} \rep{\otrans(\anedge)})^k
= 0$.

Now suppose that the condition (3) in the definition of ILECs does not hold,
i.e., there is no path from state $\astate'$ to state $\astate''$ in the graph induced by $\ilecstates$,
where all states along the path (including $\astate'$ and $\astate''$) 
belong to $\ilecstates$.
Given that $\ostates$ is finite, from some point onwards, 
any path $\apath$ features only those states from $\infstatesalong{\apath}$.
Then for any occurrence of $\astate'$ in the suffix of $\apath$ 
that features only states from $\infstatesalong{\apath}$,
if we have $\infstatesalong{\apath} = \ilecstates$,
there cannot be a subsequent occurrence of $\astate''$ along the path.
Hence we must have $\infstatesalong{\apath} \neq \ilecstates$,
and thus $\mprob{\asched}{\astate}(\set{ \apath \mid \infstatesalong{\apath} = \ilecstates}) = 0$.
\qed

\subsubsection{Proof of \lemref{lem:remain_ilec}.}
As mentioned in Section~\ref{sec:imdp_algos},
the intuition of the proof is that we define a scheduler $\asched \in \scheds{\imdp{\anoimc}}$ that 
depends only on the current state and the number of transitions done so far,
that assigns progressively decreasing probability to all edges in $\edgespara{\lz}$ that leave $\ilecstates$.
To guarantee that all states in $\ilecstates$ are visited infinitely often,
we also define a constant probability that is assigned to states in $\edgespara{\lz}$ that remain in $\ilecstates$.

In the following, we assume that there is at least one edge $(\astate',\astate'') \in \edges$
such that $\astate',\astate'' \in \ilecstates$ and $\otrans(\astate',\astate'')$ is non-singular
(otherwise the graph $(\ilecstates, \edges \cap (\ilecstates \times \ilecstates))$ is a bottom strongly connected component,
i.e., there are no edges leaving $\ilecstates$,
with fixed probabilities on each edge,
and hence $\mprob{\asched}{\astate}(\set{ \apath \mid 
\apath \not\in \reachpaths{\ostates \setminus \ilecstates} \wedge \infstatesalong{\apath} = \ilecstates}) = 1$).
Similarly, we assume that there is at least one state $\astate' \in \ilecstates$ for which
$\edgesfromthree{\lz}{\astate'}{\ostates \setminus \ilecstates}$ is non-empty;
if this is not the case, from the definition of ILECs, there is no edge that leaves $\ilecstates$.
It then remains to define the scheduler $\asched$ as selecting only assignments that 
assign fixed, positive probability to edges in $\edgespara{\lz}$ 
that have source states in $\ilecstates$
(note that these edges remain in $\ilecstates$)
in a similar manner as in the following proof,
and we omit the details.

First, we define the constant probability that is assigned by the scheduler $\asched$ to edges 
in $\edgespara{\lz}$ that have source and target states in $\ilecstates$.
For $\astate' \in \ilecstates$,
we let:
\[
\lbs{\astate'} = 
\min \set{
\min_{\anedge \in \edgesfromthree{\lz}{\astate'}{\ilecstates}} \rep{\otrans(\anedge)} \, , \,
1 - \sum_{\anedge \in \edgesfromtwo{\astate'}{\ilecstates}} \lep{\otrans(\anedge)}
} \; .
\]
Intuitively, $\lbs{\astate'}$ is the supremum probability that can be assigned to edges with left endpoint $0$
from state $\astate'$ that remain in $\ilecstates$.
We let $\lbsilec =  \frac{1}{2}\min_{\astate' \in \ilecstates} \lbs{\astate'}$,
which will be the probability assigned to edges in $\edgesfromthree{\lz}{\astate'}{\ilecstates}$
(if it is non-empty)
by the distribution chosen by $\asched$ from paths ending in state $\astate'$.

Next, for $\astate' \in \ilecstates$,
we let:
\[
\lbl{\astate'} = 
\min \set{
\min_{\anedge \in \edgesfromthree{\lz}{\astate'}{\ostates \setminus \ilecstates}} \rep{\otrans(\anedge)} \, , \,
1 - \lbsilec - \sum_{\anedge \in \edgesfromtwo{\astate'}{\ilecstates}} \lep{\otrans(\anedge)}
} \; .
\]
Intuitively, $\lbl{\astate'}$ is the supremum probability that can be assigned to edges with left endpoint $0$
from state $\astate'$ to a state \emph{not} $\ilecstates$,
assuming that probability $\lbsilec$ is assigned to edges with left endpoint $0$ that remain in $\ilecstates$
(note that this assumption is conservative, because in some states there may be no outgoing edges 
with left endpoint $0$ that remain in $\ilecstates$).
Let $\lblilec = \frac{1}{2}\min_{\astate' \in \ilecstates} \lbl{\astate'}$,
and let $\startindex$ equal the minimum index $i \geq 2$ such that $\frac{1}{2^i} < \lblilec$.

Given $\astate' \in \ilecstates$ and $i \in \Nset$,
let $\apatwo{\astate'}{i}$ be an assignment for $\astate'$ defined in the following way:
\begin{itemize}
\item
If $\edgesfromthree{\lz}{\astate'}{\ostates \setminus \ilecstates} \neq \emptyset$, 
then $\apatwo{\astate'}{i}(\astate'') = 
\frac{1}{2^{\startindex+i}} \cdot \frac{1}{|\edgesfromthree{\lz}{\astate'}{\ostates \setminus\ilecstates}|}$ 
for each $\astate'' \in \ostates \setminus \ilecstates$ such that 
$(\astate',\astate'') \in \edgesfromthree{\lz}{\astate'}{\ostates \setminus\ilecstates}$.
\item
If $\edgesfromthree{\lz}{\astate'}{\ilecstates} \neq \emptyset$, 
then  
$\apatwo{\astate'}{i}(\astate'') = \lbsilec \cdot \frac{1}{|\edgesfromthree{\lz}{\astate'}{\ilecstates}|}$ 
for each $\astate'' \in \ilecstates$ such that $(\astate',\astate'') \in \edgesfromthree{\lz}{\astate'}{\ilecstates}$.
\end{itemize}
The definitions of $\lbsilec$ and $\lblilec$ allow us to complete the definition of $\apatwo{\astate'}{i}$
for states not considered in these two points above,
so that it $\apatwo{\astate'}{i}$ an assignment for $\astate'$.
The intuition here is as follows:
condition~(2) of the definition of ILECs specifies that obtaining an assignment
that obeys the above constraints and which sums to $1$ is not problematic;
furthermore, the definitions of $\lbsilec$ and $\lblilec$
guarantee that enough probability mass is available to edges that do \emph{not} have 
a left endpoint of $0$ to equal or exceed their left endpoints,
and that the probability that is assigned to edges that \emph{do} have a left endpoint 
of $0$ does not exceed their right endpoint.

We now formalise the definition of $\asched$.
Let $\afinpath \in \mfinpaths{\imdp{\anoimc}}(\astate)$, 
where $\afinpath = \astate_0 \adist_0 \astate_1 \adist_1 \cdots \adist_{n-1} \astate_n$
and $\astate_i \in \ilecstates$ for all $i \leq n$.
We let $\asched(\afinpath) = \dirac{ \apatwo{\astate_n}{n} }$.

\sloppypar{
It remains to show that $\mprob{\asched}{\astate}(\set{ \apath \mid 
\apath \not\in \reachpaths{\ostates \setminus \ilecstates} \wedge \infstatesalong{\apath} = \ilecstates}) > 0$.
For a path $\apath \in \mpaths{\imdp{\anoimc}}(\astate)$,
let $\apath(i)$ be the $(i+1)$-th state along the path.
Given state set $\astateset \subseteq \ostates$ and $k \in \Nset$,
we let $\boundedreachpaths{k}{\astateset}$ be the set of paths that reach $\astateset$ within $k$ transitions;
formally, $\boundedreachpaths{k}{\astateset} = 
\set{ \apath \in \mpaths{\imdp{\anoimc}}(\astate) \mid \exists i \leq k \ldot \apath(i) \in \astateset }$.
By definition, we have $\reachpaths{\astateset} = \bigcup_{k \in \Nset} \boundedreachpaths{k}{\astateset}$,
and $\mprob{\asched}{\astate}(\reachpaths{\astateset}) = 
\lim_{k \ra \infty} \mprob{\asched}{\astate}(\boundedreachpaths{k}{\astateset})$.
From the definition of $\asched$ above,
we have $\mprob{\asched}{\astate}(\boundedreachpaths{0}{\ostates \setminus \ilecstates}) = 0$
(from $\astate \in \ilecstates$) and,
for $k \geq 1$, we have 
$\mprob{\asched}{\astate}(\boundedreachpaths{k}{\ostates \setminus \ilecstates}) 
\leq 
\frac{1}{4} + \sum_{i=1}^k \frac{1}{2^{i+2}} \prod_{j=0}^{i-1} (1-\frac{1}{2^{i+1}})
\leq
\frac{1}{4} + \sum_{i=1}^k \frac{1}{2^{i+2}}
= 
\sum_{i=0}^k \frac{1}{2^{i+2}}$.
We note that $\sum_{i=0}^k \frac{1}{2^{i+2}} \leq \frac{1}{2}$ for all $k \in \Nset$.
Hence $\mprob{\asched}{\astate}(\reachpaths{\ostates \setminus \ilecstates}) = 
\lim_{k \ra \infty} \mprob{\asched}{\astate}(\boundedreachpaths{k}{\ostates \setminus \ilecstates})
\leq 
\lim_{k \ra \infty} \sum_{i=0}^k \frac{1}{2^{i+2}}
\leq \frac{1}{2}$.
Therefore we have shown that $\mprob{\asched}{\astate}(\set{ \apath \mid 
\apath \not\in \reachpaths{\ostates \setminus \ilecstates}}) \geq \frac{1}{2} > 0$.
}

Because the assignments used by $\asched$ dedicate a probability value to all edges 
with source and target states in $\ilecstates$ that is no lower than some fixed lower bound,
and because the graph $(\ilecstates, \edges \cap (\ilecstates \times \ilecstates))$ is strongly connected,
we have that $\mprob{\asched}{\astate}(\set{ \apath \mid 
\apath \not\in \reachpaths{\ostates \setminus \ilecstates} \wedge \infstatesalong{\apath} = \ilecstates})
= \mprob{\asched}{\astate}(\set{ \apath \mid 
\apath \not\in \reachpaths{\ostates \setminus \ilecstates}})$.
Hence we have shown that $\mprob{\asched}{\astate}(\set{ \apath \mid 
\apath \not\in \reachpaths{\ostates \setminus \ilecstates} \wedge \infstatesalong{\apath} = \ilecstates}) > 0$.
\qed 

\begin{remark}
Note that we can confine the IMC to an ILEC with probability $1$ 
if the only edges leaving the ILEC belong to $\edgespara{\lcz}$; 
however, we do not require that result,
and above we settle for a simplified scheduler construction that does not distinguish between 
edges in $\edgespara{\lcz}$ and $\edgespara{\loz}$,
and hence allows some progressively decreasing probability of exiting from the ILEC
even if the outgoing edges of the ILEC are only from $\edgespara{\lcz}$.
\end{remark}

\subsubsection{Proof of \propref{prop:imdp_ltone}.}
$(\Rightarrow)$
Let $\asched \in \scheds{\imdp{\anoimc}}$ be 
such that $\mprob{\asched}{\astate}(\reachpaths{\targetset}) < 1$.
Then, by duality, we have $\mprob{\asched}{\astate}(\set{ \apath \mid 
\apath \not\in \reachpaths{\targetset}}) > 0$.
From this fact, and from \lemref{lem:fund_ilecs},
we then have that $\mprob{\asched}{\astate}(\set{ \apath \mid 
\apath \not\in \reachpaths{\targetset} \wedge \infstatesalong{\apath} \in \ilecs}) > 0$.
For any path $\apath \in \mpaths{\asched}(\astate)$,
we have that $\apath \not\in \reachpaths{\targetset}$ and $\infstatesalong{\apath} \in \ilecs$
implies that $\apath \in \reachpaths{\statesettwo{\ilecs}{\targetset}}$
from the definition of $\statesettwo{\ilecs}{\targetset}$.
Hence $\mprob{\asched}{\astate}(\set{ \apath \mid 
\apath \in \reachpaths{\statesettwo{\ilecs}{\targetset}}}) > 0$.
From \lemref{lem:reach_imc},
we then have that there exists a finite path $\afinpath \in \gfinpaths{\anoimc}(\astate)$ 
such that $\last(\afinpath) \in \statesettwo{\ilecs}{\targetset}$,
completing this direction of the proof.

$(\Leftarrow)$
The existence of a path $\afinpath \in \gfinpaths{\anoimc}(\astate)$ 
such that $\last(\afinpath) \in \statesettwo{\ilecs}{\targetset}$
implies that there exists $\asched \in \scheds{\imdp{\anoimc}}$
such that $\mprob{\asched}{\astate}(\set{ \apath \mid 
\apath \in \reachpaths{\statesettwo{\ilecs}{\targetset}}}) > 0$
by \lemref{lem:reach_imc}.
We define $\asched' \in \scheds{\imdp{\anoimc}}$
such that $\mprob{\asched'}{\astate}(\reachpaths{\targetset}) < 1$ in the following way:
the scheduler $\asched'$ behaves as $\asched$ until a state in $\statesettwo{\ilecs}{\targetset}$ is reached;
once such a state has been reached, 
the scheduler then behaves as a scheduler defined in \lemref{lem:remain_ilec},
which ensures that the states ILEC that does not contain any state from $\targetset$
are henceforth visited infinitely often with positive probability.
The construction of the scheduler is standard,
and we describe it here for completeness.

Let $\astate' \in \statesettwo{\ilecs}{\targetset}$,
and let $\ilecstatespara{\astate'} \in \ilecs$ be the maximal ILEC such that
$\astate' \in \ilecstatespara{\astate'}$ and $\ilecstatespara{\astate'} \cap \targetset = \emptyset$
(which exists by definition of $\statesettwo{\ilecs}{\targetset}$).
Now let $\aschedpara{\astate'} \in \scheds{\imdp{\anoimc}}$ be such that 
$\mprob{\asched}{\astate'}(\set{ \apath \mid 
\apath \not\in \reachpaths{\ostates \setminus \ilecstatespara{\astate'}} \wedge 
\infstatesalong{\apath} = \ilecstatespara{\astate'}}) > 0$,
which exists by \lemref{lem:remain_ilec}.
For finite paths $\afinpath = 
\astate_0 \adist_0 \astate_1 \adist_1 \cdots \adist_{n-1} \astate_n \in \mpaths{\imdp{\anoimc}}(\astate)$
such that $\astate_i \not\in \statesettwo{\ilecs}{\targetset}$ for all $i \leq n$,
we let $\asched'(\afinpath) = \asched(\afinpath)$.
For a finite path $\afinpath = 
\astate_0 \adist_0 \astate_1 \adist_1 \cdots \adist_{n-1} \astate_n \in \mpaths{\imdp{\anoimc}}(\astate)$
such that there exists $i \leq n$ for which 
$\astate_j \not\in \statesettwo{\ilecs}{\targetset}$ for all $j < i$
and $\astate_j \in \statesettwo{\ilecs}{\targetset}$ for all $j \geq i$,
we let $\asched'(\afinpath) = \aschedpara{\astate_i}(\astate_i \adist_i \cdots \adist_{n-1} \astate_n)$.
For other finite paths, the definition of $\asched'$ can be arbitrary.
Given that $\mprob{\asched}{\astate}(\set{ \apath \mid 
\apath \in \reachpaths{\statesettwo{\ilecs}{\targetset}}}) > 0$,
we have $\mprob{\asched'}{\astate}(\set{ \apath \mid 
\apath \in \reachpaths{\statesettwo{\ilecs}{\targetset}}}) > 0$,
and we know that there exists a finite path $\afinpath \in \mfinpaths{\asched'}(\astate)$ 
(with positive probability)
such that $\last(\afinpath) \in \statesettwo{\ilecs}{\targetset}$.
Given the definition of the behaviour of $\asched'$ after finite path $\afinpath$,
we have $\mprob{\asched'}{\astate}(\set{ \apath \mid 
\afinpath \mbox{ is a prefix of } \apath \wedge
\apath \not\in \reachpaths{\ostates \setminus \ilecstatespara{\last(\afinpath)}} \wedge 
\infstatesalong{\apath} = \ilecstatespara{\last(\afinpath)}}) > 0$.
This then implies that $\mprob{\asched'}{\astate}(\set{ \apath \mid 
\apath \not\in \reachpaths{\ostates \setminus \ilecstatespara{\last(\afinpath)}} \wedge 
\infstatesalong{\apath} = \ilecstatespara{\last(\afinpath)}}) > 0$.
Because we have asumed that all states in $\targetset$ are absorbing,
the set $\targetset$ is not reached along $\afinpath$
(otherwise $\statesettwo{\ilecs}{\targetset}$ could not be the final state of $\afinpath$);
then, given that $\ilecstatespara{\last(\afinpath)} \cap \targetset = \emptyset$,
we have that $\mprob{\asched'}{\astate}(\set{ \apath \mid 
\apath \not\in \reachpaths{\targetset}}) > 0$,
and by duality $\mprob{\asched'}{\astate}(\reachpaths{\targetset}) < 1$.
Hence this direction of the proof is completed.
\qed

\end{document}